\documentclass[preprint,showpacs,preprintnumbers,amsmath,amssymb,aps]{revtex4}

\usepackage{graphicx}
\usepackage{dcolumn}
\usepackage{bm}

\begin{document}

\title{Hyperon polarization in different inclusive production 
processes in unpolarized high energy hadron-hadron collisions}
\author{Dong Hui}\email{donghui@mail.sdu.edu.cn}
\author{Liang Zuo-tang}\email{liang@sdu.edu.cn}
\affiliation{Department of Physics, Shandong University, 
Jinan, Shandong 250100, China}

\begin{abstract}
We apply the picture proposed in a previous Letter, 
which relates the hyperon polarization 
in unpolarized hadron-hadron collisions 
to the left-right asymmetry in singly polarized reactions, 
to the production of different hyperons 
in reactions using different projectiles and/or targets. 
We discuss the different ingredients of the proposed picture in detail 
and present the results for hyperon polarization 
in the reactions such as $pp$, $K^-p$, $\pi^{\pm}p$, and $\Sigma^-p$ collisions. 
We compare the results with the available data 
and make predictions for future experiments. 
\end{abstract}

\pacs{13.88.+e, 13.85.Ni, 13.85.-t}
\maketitle

\section{introduction}
\label{sec:introduction}
Since the discovery \cite{Les75,Bun76} in 1970s, 
the surprisingly large transverse polarization of hyperons 
in unpolarized high energy hadron-hadron and hadron-nucleus collisions 
has been a standing hot topic in High Energy Spin 
Physics (see e.g, \cite{Les75,Bun76,Hel96,And79,Deg81,Szw81,Pon85,Bar92,Sof92,LB97,LB00,LLts00,Ans01}, 
and the references cited therein.).
Experimentally, there are a large number of similar experiments 
that have been performed 
at different energies and/or using different projectiles 
and/or targets and for the production of different hyperons \cite{Hel96}. 
Theoretically, different models 
have been proposed \cite{And79,Deg81,Szw81,Pon85,Bar92,Sof92,LB97,LB00,LLts00,Ans01},  
the aim of which is to understand the origin(s) 
of this striking spin effect in high energy reactions. 
Clearly such studies should provide us 
with useful information on the spin structure of the hadron 
and the spin dependence of strong interactions. 

Inspired by the similarities of the corresponding 
data \cite{Hel96,ANdata,E704}, 
we proposed a new approach in a recent Letter \cite{LB97} 
to understand the origin(s) of the transverse hyperon polarization 
in unpolarized hadron-hadron collisions 
by relating them to the left-right asymmetries 
observed \cite{ANdata,E704} in singly polarized $pp$ collisions. 
We pointed out that these two striking spin phenomena 
should be closely related to each other 
and have the same origin(s). 
We showed that using the spin correlation deduced 
from the single-spin left-right asymmetries 
for inclusive $\pi$ production as input, 
we can naturally understand the transverse polarization 
for a hyperon which has one valence quark 
in common with the projectile, such as $\Sigma^-$, 
$\Xi^0$, or $\Xi^-$ in $pp$ collisions, 
or $\Lambda$ in $K^-p$ collisions.
We showed also that, 
to understand the puzzling transverse polarization 
of $\Lambda$ in $pp$ collisions, 
which have two valence quarks in common with the projectile, 
we need to assume that the $s$ and $\bar{s}$, 
which combine respectively 
with the valence-($ud$) diquark and the remaining $u$-valence quark 
to form the produced $\Lambda$ and the associatively produced $K^+$ 
in the fragmentation region, 
should have opposite spins. 
Under this assumption, we obtained a good quantitative fit 
to the $x_F$ dependence of $\Lambda$ polarization in $pp$ collisions 
(where $x_F\equiv 2p_\|/\sqrt{s}$, $p_\|$ is the longitudinal component 
of the momentum of the produced hyperon, and $\sqrt{s}$ 
is the total center of mass energy of the $pp$ system). 
The obtained qualitative features for the polarizations 
of other hyperons are all in good agreement with the available data. 

There are two main points 
in the picture that need to be further tested, 
i.e., (i) the $s$ and $\bar{s}$ 
that combine respectively 
with the valence-($ud$) diquark and the remaining $u$-valence quark 
of the projectile proton to form the produced $\Lambda$ 
and the associatively produced $K^+$ 
in the fragmentation region
have opposite spins, 
and (ii) the SU(6) wave function 
can be used to describe the relation between the spin of 
the fragmenting quark and that of the hadron 
produced in the fragmentation process. 
Developments have been made since the publication of Ref. \cite{LB97}. 
We found out that exclusive reactions 
such as $pp\to p\Lambda K^+$ and $e^-p\to e^-\Lambda K^+$ 
are best suitable to test point (i). 
We therefore applied the picture to these processes 
and presented the obtained results in Refs. \cite{LB00} and \cite{LX02}. 
It is encouraging to see that these results 
are all in agreement with the available data \cite{R608,CLAS03}. 

It has also been pointed out \cite{BL98} 
that the longitudinal polarization of $\Lambda$ 
in $e^+e^-$ annihilation at $Z^0$ pole 
provides a special test to point (ii), 
i.e., whether SU(6) wave function can be used 
in relating the spin of the fragmenting quark 
to that of the produced hadron. 
Calculations have been made \cite{BL98,GH93} 
and the obtained results are consistent with the data \cite{Aleph96,OPAL98}. 
Since neither the accuracy nor the abundance of the data is high enough 
to give a conclusive judgment, 
we made a systematic study \cite{BL98,LL00} of hyperon polarization 
in other reactions which can be used 
to test this point. 
The results are in agreement with 
the available data \cite{Aleph96,OPAL98,NOMAD00} 
and future experiments are under way. 

Encouraged by these developments, 
in this paper, we apply the proposed picture 
to study the polarizations of different hyperons 
in different unpolarized hadron-hadron and hadron-nucleus reactions. 
We summarize the different ingredients of the picture 
and their developments in detail 
in Sec. \ref{sec:picture}. 
In Sec. \ref{sec:method}, 
we give the calculation method 
of hyperon polarization 
in different processes using the picture. 
In Sec. \ref{sec:results}, 
we apply the method to calculate the polarizations 
of different hyperons produced in unpolarized $pp$, 
$K^-p$, $\pi^\pm p$, and $\Sigma^-p$ collisions. 
We present the results obtained 
and compare them with the available data. 
Finally, a short summary and outlook is given in Sec. V.

\section{the physical picture}
\label{sec:picture}
In this section, we summarize 
the key points of the picture proposed in Ref. \cite{LB97}. 
The basic idea of the picture 
is that there should be a close relation 
between hyperon polarization ($P_H$) 
in unpolarized hadron-hadron collisions 
and the left-right asymmetry ($A_N$) 
in single-spin hadron-hadron collisions. 
Hence, if we extract the essential information encoded 
in the $A_N$ data, 
we can study $P_H$ based on such information. 
There are three key points in this physical picture 
which are summarized as the following.

\subsection{Correlation between the spin of the quark 
and the direction of motion of the produced hadron}
\label{subsec:a}
It has been pointed out \cite{LB97} that the existence of $A_N$ 
in singly polarized hadron-hadron collision 
implies the existence of a spin correlation 
between the spin of the fragmenting quark 
and the direction of momentum of the produced hadron, 
i.e., $\vec{s}_q\cdot\vec{n}$ type of spin correlation 
in the reaction. 
[Here, $\vec{s}_q$ is the spin of the quark; 
$\vec{n}\equiv(\vec{p}_{inc}\times\vec{p}_h)/|\vec{p}_{inc}\times\vec{p}_h|$ 
is the unit vector in the normal direction of the production plane, 
$\vec{p}_{inc}$ and $\vec{p}_h$ are respectively the momentum 
of the incident hadron and that of the produced hadron.] 
One of the major ingredients of the picture proposed in Ref. \cite{LB97} 
is that both the existence of $A_N$ and that of $P_H$ 
are different manifestations of this spin correlation 
$\vec{s}_q\cdot\vec{n}$. 
Hence we can use the experimental results for $A_N$ 
as input to determine the strength of this spin correlation, 
then apply it to unpolarized hadron-hadron collision 
to study $P_H$. 

We recall that \cite{ANdata,E704,LB00rev}, in the language commonly used 
in describing $A_N$, 
the polarization direction of the incident proton is called upward, 
and the incident direction is forward. 
The single-spin left-right asymmetry $A_N$ is just the difference 
between the cross section where $\vec{p}_h$ points 
to the left and that to the right, 
which corresponds to $\vec{s}_q\cdot\vec{n}=1/2$ 
and $\vec{s}_q\cdot\vec{n}=-1/2$, 
respectively. 
The data \cite{ANdata,E704} on $A_N$ show that 
if a hadron is produced by an upward polarized 
valence quark of the projectile, 
it has a large probability to have a transverse momentum 
pointing to the left. 
$A_N$ measures the excess of hadrons 
produced to the left over those produced to the right. 
The difference of the probability for the hadron 
to go left and that to go right 
is denoted \cite{BLM93,LB00rev} by $C$ 
if the hadron is produced by an upward polarized quark. 
$C$ is a constant in the range of $0<C<1$. 
It has been shown that \cite{BLM93,LB00rev}, 
to fit the $A_N$ data \cite{E704} 
in the transverse momentum interval $0.7<p_T<2.0$ GeV/$c$, 
$C$ should be taken as $C=0.6$. 

In terms of the spin correlation discussed above, 
the cross section should be expressed as 
\begin{equation}
\label{eq:sigma}
\sigma=\sigma_0+(\vec{s}_q\cdot\vec{n})\sigma_1, 
\end{equation}
where $\sigma_0$ and $\sigma_1$ are 
independent of $\vec{s}_q$. 
The second term just denotes 
the existence of the $\vec{s}_q\cdot\vec{n}$ type 
of spin correlation. 
$C$ is just the difference 
between the cross section where $\vec{s}_q\cdot\vec{n}=1/2$ 
and that where $\vec{s}_q\cdot\vec{n}=-1/2$ 
divided by the sum of them, 
i.e., $C=\sigma_1/(2\sigma_0)$. 

Now, we assume the same strength for the spin correlation 
in hyperon production in the same collisions. 
It follows that the quark 
which fragments into the hyperon should be polarized, 
and the polarization $P_q$ can be determined 
by using Eq. (\ref{eq:sigma}). 
Since both the left-right asymmetry in singly polarized collision 
and hyperon polarization in unpolarized collision 
exist mainly in large $x_F$ region, 
we assume that the spin correlation exists 
only for valence quarks of the incident hadrons. 
For a hyperon produced with momentum $\vec{p}_h$, 
$\vec{n}$ is given. 
The cross section that this hyperon is produced 
in the fragmentation of a valence quark 
with spin satisfying $\vec{s}_q\cdot\vec{n}=1/2$ 
is $(\sigma_0+\sigma_1/2)$, 
and that with spin satisfying $\vec{s}_q\cdot\vec{n}=-1/2$ 
is $(\sigma_0-\sigma_1/2)$. 
Hence, the polarization of the valence quarks 
which lead to the production of the hyperons 
with that $\vec{n}$ is given by, 
\begin{equation}
\label{eq:P_q}
P_q=\frac{(\sigma_0+\sigma_1/2)-(\sigma_0-\sigma_1/2)}
{(\sigma_0+\sigma_1/2)+(\sigma_0-\sigma_1/2)}
=\frac{\sigma_1}{2\sigma_0}=C. 
\end{equation}

It should be emphasized 
that $P_q\ne 0$ just means that the strength 
of the spin correlation of the form $\vec{s}_q\cdot\vec{n}$ 
is nonzero in the reaction. 
It means that, due to some spin-dependent interactions, 
the quarks which have spins along the same direction 
as the normal of the production plane 
have a large probability to combine with suitable 
sea quarks to form the specified hyperons than those 
which have spins in the opposite direction. 
It does not imply that the quarks 
in the unpolarized incident hadrons 
were polarized in a given direction, 
which would contradict the general requirement 
of space rotation invariance. 
In fact, in an unpolarized reaction, 
the normal of the production plane of the specified hyperons 
is uniformly distributed in the transverse directions. 
Hence, averaging over all the normal directions, 
the quarks are unpolarized. 

We would like also to mention that similar idea 
has been applied \cite{XL03} to spin alignments of vector mesons 
in unpolarized hadron-hadron collisions.  
It has been shown that the existence of the spin alignment 
of vector mesons in unpolarized hadron-hadron collision 
is another manifestation of the existence of the 
$\vec s_q\cdot \vec n$ type of spin correlation. 
The obtained result are in agreement with the 
available data \cite{Chl72,Bar83,EXCHARM00}.

\subsection{Relating the spin of the quark to that of the hadron}
\label{subsec:b}
As discussed in the above-mentioned subsection, 
the existence of the $\vec{s}_q\cdot\vec{n}$ type 
of spin correlation in hadron-hadron collision 
implies a polarization of the quark $q^0$ 
transverse to the production plane of the hyperon. 
(Here, we use the superscript $0$ to denote the quark before fragmentation.) 
To study the polarization of the produced hyperon
from this point, 
we need to know the relation between the spin 
of the quark and that of the hadron produced 
in the fragmentation of this quark. 
The question of the relation 
between the spin of the fragmenting quark $q^0$ 
and that of the hadron created in the fragmentation 
of $q^0$ is usually referred to 
as ``spin transfer in high energy fragmentation process''. 
It contains two parts: 
will $q^0$ keep its polarization in the fragmentation? 
what is the relation between the spin of $q^0$ 
and that of the hadron which contains $q^0$? 

The answers to these questions depend 
on the spin structure of hadron 
and the hadronization mechanism. 
They can even be different in the longitudinally polarized case 
from those for the transversely polarized case. 
Since neither of them can be solved 
using perturbative calculations, 
presently, phenomenological studies are need 
to search the answers to them. 
Currently, there exist two distinct pictures 
for the spin structure of nucleon, 
i.e., the SU(6) picture 
based on the SU(6) wave function of the baryon, 
and the DIS picture 
based on the polarized deeply inelastic lepton-nucleon 
scattering data and other inputs such as symmetry assumptions 
and data from other experiments. 
It is of particular interest to know 
which one is suitable here. 

It is clear that to study these questions, 
one needs to know the polarization 
of the quark before fragmentation 
and measure the polarization of the hadron 
produced in the fragmentation. 
Hence, we have the following 
two possibilities: 
One is to study hyperon polarization in 
$e^+e^-$ annihilation at $Z^0$ pole, 
in polarized $ep$ deeply inelastic scattering 
or in high $p_T$ polarized $pp$ collision.
The other is to study the vector meson polarization 
in these processes.

It has been pointed out 
that the $\Lambda$ polarization in $e^+e^-$ annihilation 
at the $Z^0$ pole provides a very special test 
to the applicability of the SU(6) picture 
in the longitudinally polarized case. 
This is because, 
for $e^+e^-\to Z^0\to \Lambda X$, 
the $s\bar{s}$ created 
at the $e^+e^-$ annihilation vertex 
is almost completely longitudinally polarized. 
In this case, if we assume 
that the quark keeps its polarization in the fragmentation 
and use the SU(6) wave function 
to connect the spin of the quark 
and that of the hyperon that contains this quark, 
we should obtain a maximum for the magnitude 
of $\Lambda$ polarization 
since in this picture the spin of the $s$-quark 
is completely transferred to $\Lambda$. 
Experimental data were obtained \cite{Aleph96,OPAL98} 
for $e^+e^-\to Z^0\to\Lambda X$ 
by the ALEPH and OPAL Collaborations at LEP. 
We showed \cite{BL98} that if we assume 
that the $q^0$ keeps its polarization in fragmentation 
and the SU(6) wave function can be used 
in relating the polarization of $q^0$ 
and that of the produced hyperon which contains $q^0$, 
we obtain the results which are 
in agreement with the data. 
This result is rather encouraging. 
But, the accuracy and abundance of the data 
are not enough to make a conclusive judgment. 
In particular, there is no direct measurement 
available at all in the transversely polarized case. 
We therefore made a systematic calculation 
for hyperon polarizations 
in all the different reactions \cite{LL00}. 
There are also data for $\Lambda$ polarization 
in deeply inelastic scattering \cite{NOMAD00}, 
they are also consistent with the results 
obtained using the SU(6) picture.

There are also data 
that provide information 
on vector meson polarization 
in high energy reactions. 
The $00$-elements 
of the helicity density matrices 
for $K^*$, $\rho$, etc in $e^+e^-\to Z^0\to VX$ 
have been measured \cite{ALEPH95,DELPHI95,OPAL97} 
by the ALEPH, DELPHI and OPAL Collaborations at LEP. 
We showed \cite{XLL01} that these data can also 
be understood using the SU(6) picture. 
Further tests, not only in the longitudinally polarized case 
but also in the transversely polarized case, are under way. 
In this paper, we assume it is the same 
in longitudinally and transversely polarized cases 
and use it in studying hyperon polarization 
in unpolarized high energy hadron-hadron collisions. 

Having the two points discussed in the last and this subsections, 
we can already obtain the polarizations 
for those hyperons which have one valence quark 
of the same flavor as that of the projectile, 
e.g., $pp\to\Sigma^-X$, $K^-p\to\Lambda X$ 
and $\Sigma^-p\to\Sigma^+X$. 
Some of the qualitative features of the results 
are given in Ref. \cite{LB97}. 
It is encouraging to see that all of them 
are in good agreement with the data \cite{Hel96}. 

\subsection{Correlation between the spin of 
$q_s$ and that of $\bar{q}_s$ which combine with 
the $(q_vq_v)$ and remaining $q_v$ 
to form the produced hyperon and the associated meson}
\label{subsec:c}
To study the polarization of hyperons 
such as $\Lambda$ in $pp\to\Lambda X$, 
i.e., those which have two valence quarks 
that have the same flavors as those of the projectile, 
we encounter the following question: 
If a hadron is produced 
by two valence quarks (valence diquark) $q_vq_v$ of the projectile, 
the remaining valence quark $q_v$ produces an associated hadron. 
What are the spin states of 
the $q_s$ and $\bar{q}_s$ 
that combine with the $q_vq_v$ and the remaining $q_v$ 
to form the produced hyperon and the associatively produced meson 
in the fragmentation region, respectively? 

Theoretically, it is quite difficult to derive it 
since we are in the very small $x$ region; 
the production of such pairs 
is also of soft nature in general 
and cannot be calculated using perturbative theory. 
To get some clue to this problem, 
we still start from the single-spin left-right 
asymmetry $A_N$. 
The existing data \cite{ANdata,E704} clearly show that 
$A_N$ in $p^\uparrow p\to \Lambda X$ 
is large in magnitude and negative in sign 
in the fragmentation region. 
We note that the $\Lambda$ in this region 
is mainly produced 
by the valence-$(u_vd_v)$-diquark of the projectile 
and is associated with the production 
of a $K^+$ produced by the remaining $u_v$. 
From the SU(6) wave function 
we learn that the $(u_vd_v)$ has to be 
in the spin zero state 
and the spin of the proton 
is carried by the remaining $u_v$. 
According to the $\vec{s}_q\cdot\vec{n}$ 
spin correlation discussed in Subsection \ref{subsec:a}, 
$A_N$ for the associatively produced $K^+$ is positive. 
Hence, to understand the data on $A_N$ for $\Lambda$, 
we simply need to assume \cite{BL96} 
that $\Lambda$ produced by $(u_vd_v)$ 
and $K^+$ that is associatively produced 
by the remaining $u_v$ 
move in the opposite transverse directions, 
which is just a direct consequence 
of transverse momentum conservation. 
Now, we apply this to unpolarized $pp$ collision 
and consider the case 
that a $\Lambda$ is produced by $(u_vd_v)$ 
together with a $s$-quark 
and a $K^+$ is associatively produced 
by the remaining $u_v$ together with the $\bar{s}$. 
According to the $\vec{s}_q\cdot\vec{n}$ 
spin correlation mentioned in Subsection \ref{subsec:a}, 
the remaining $u_v$ should have a large probability 
to be polarized in $-\vec{n}$ direction 
since the normal of the production plane for the $K^+$ 
is opposite to the normal $\vec{n}$ 
of the production plane of $\Lambda$. 
The polarization is $-C$. 
Since $K^+$ is a spin zero object, 
the $\bar{s}$ should have 
a polarization of $+C$ 
(in the $\vec{n}$ direction). 
The data for $\Lambda$ polarization 
show that $P_\Lambda$ is large and negative. 
This means that the $s$-quark 
has a negative polarization. 
We thus reach the conclusion 
that, to understand the polarization of $\Lambda$ in 
$pp\to \Lambda X$, we need to assume that  
the $s$ and $\bar{s}$ have opposite spins. 
Under this assumption, together with 
the two points mentioned in the last two subsections, 
we obtained \cite{LB97} a good fit to the data on $\Lambda$ polarization. 

This point needs of course to be further studied 
and tested experimentally. 
We found out that the simple exclusive process 
are most suitable for this purpose. 
Hyperon polarizations 
in the exclusive processes such as 
$pp\to p\Lambda K^+$ and $e^-p\to e^-\Lambda K^+$ 
are very sensitive to the spin states 
of the $s$ and $\bar{s}$ pairs. 
If the $s$ and $\bar{s}$ have opposite spins, 
the obtained results 
for $\Lambda$ polarization in $pp\to p\Lambda K^+$ 
should take the maximum among the different channels 
for $pp\to\Lambda X$. 
This is because that here we have a situation 
that the $\Lambda$ is definitely produced by the 
$(u_vd_v)$ valence diquark and is definitely 
associated with a $K^+$ that is definitely 
produced by the remaining $u_v$ of 
the incident $p$. 
Hence, we obtain that in this case, 
$P_\Lambda=-C=-0.6$. 
This is in good agreement with the data obtained \cite{R608} 
by the R608 Collaboration at CERN
which show that $P_\Lambda=-0.62\pm 0.04$. 

We also calculated \cite{LX02} $\Lambda$ polarization 
in $e^-p\to e^-\Lambda K^+$ 
in all three cases in which the spin states 
of $s$ and $\bar{s}$ can be, 
i.e., opposite, same or uncorrelated. 
We found out that the results in the three cases 
are quite different from each other. 
Now, experimental data are obtained \cite{CLAS03} 
for $\Lambda$ polarization in $e^-p\to e^-\Lambda K^+$
by the CLAS Collaboration at Jefferson Laboratory. 
Comparing to the data, 
we see that the results obtained in the case 
that the spins of $s$ and $\bar{s}$ are opposite 
are favored. 
Further tests are also under way. 

We now assume that this is in general true, 
i.e., the $q_s$ that combines with the $(q_vq_v)$ 
of the projectile to form the hyperon 
and the $\bar{q}_s$ that combines with the remaining $q_v$ 
to form the associatively produced meson 
have opposite spins. 
Under this assumption, we obtain the result that 
the $q_s$ should be polarized in the $-\vec n$ direction 
and the polarization is $-C$.
We apply this to the production of different hyperons  
to calculate the hyperon polarizations 
in unpolarized high energy hadron-hadron or 
hadron-nucleus collisions in next sections. 

We emphasize that the above mentioned result is true 
for the production of hyperons associated with the 
production of pseudo-scalar mesons. 
The situation should be different if the 
associatively produced meson is a vector meson.
This influence will be further investigated in a separate paper \cite{DLfuture} and 
here we consider only the former case.

\section{the calculation method}
\label{sec:method}
Having the picture discussed in last section, 
we can calculate the hyperon polarization 
in different hadron-hadron collisions. 
We now present the formulas used in these calculations. 

\subsection{General formulas}
\label{subsec:general}
We consider the process $P+T\to H_i+X$, 
where $P$ and $T$ denote respectively 
the projectile and target hadron, 
and $H_i$ denotes the $i$th kind of the hyperons. 
The hyperon polarization $P_{H_i}$ is defined as, 
\begin{equation}
\label{eq:define}
P_{H_i}(x_F|s)\equiv\frac{N(x_F,H_i,\uparrow|s)-N(x_F,H_i,\downarrow|s)}
{N(x_F,H_i,\uparrow|s)+N(x_F,H_i,\downarrow|s)}
=\frac{\Delta N(x_F,H_i|s)}{N(x_F,H_i|s)},
\end{equation}
where $N(x_F,H_i,l|s)$ is the number density of $H_i$'s 
polarized in the same ($l=\uparrow$) 
or opposite ($l=\downarrow$) direction 
as the normal ($\vec{n}$) of the production plane 
at a given $\sqrt{s}$; 
$x_F\equiv2p_\|/\sqrt{s}$, 
$p_\|$ is the longitudinal momentum of $H_i$ 
with respect to the incident direction of $P$, 
and $\sqrt{s}$ is the total c.m. energy of the 
colliding hadron system. 
It is clear that the denominator is nothing else 
but the number density of $H_i$ without specifying the polarization.

To calculate $\Delta N(x_F,H_i|s)$, 
we divide the final hyperons $H_i$'s 
into the following four groups 
according to the different origins for the production: 
(A) those directly produced 
and contain a valence diquark (two valence quarks) 
$(q_vq_v)^P$ of the projectile; 
(B) those directly produced 
and contain a valence quark $q_v^P$ of the projectile; 
(C) those from the decay of the directly produced heavier hyperons $H_j$'s 
that contain a $(q_vq_v)^P$ or a $q_v^P$; 
and (D) the others. 
In this way, we have, 
\begin{equation}
\label{eq:N}
N(x_F,H_i|s)=N_0(x_F,H_i|s)+D^A(x_F,H_i|s)
+\sum_f{D^{B,f}(x_F,H_i|s)}+\sum_j{D^{C,H_j}(x_F,H_i|s)}, 
\end{equation}
where $D^A(x_F,H_i|s)$, $D^{B,f}(x_F,H_i|s)$ 
and $D^{C,H_j}(x_F,H_i|s)$ 
denote the contributions from groups (A), (B) and (C), respectively; 
the superscript $f$ and $j$ 
denote the flavor of $q_v$ and the type of $H_j$, respectively; 
$N_0$ is the contribution from (D). 

According to the picture discussed in last section, 
hyperons from groups (A), (B) and (C) 
can be polarized, 
while those from (D) are not. 
This means that those from (A), (B) and (C) 
contribute to the numerator $\Delta N$ of Eq. (\ref{eq:N}), 
i.e., we have, 
\begin{equation}
\label{eq:DeltaN}
\Delta N(x_F,H_i|s)=\Delta D^A(x_F,H_i|s)
+\sum_f{\Delta D^{B,f}(x_F,H_i|s)}+\sum_j{\Delta D^{C,H_j}(x_F,H_i|s)}. 
\end{equation}

Since valence quarks usually carry large fractions of 
the momenta of the incident hadrons, 
we expect that, for very large $x_F$, 
$D^A(x_F,H_i|s)$ dominates. 
For small $x_F$, $N_0$ dominates, 
while for moderate $x_F$, $D^B(x_F,H_i|s)$ 
plays the dominant role. 
Hence, if we neglect the decay contributions, 
we expect from Eqs. (\ref{eq:define}--\ref{eq:DeltaN}) 
that $P_{H_i}(x_F|s)$ has the following general properties. 
For $x_F$ increasing from 0 to 1, 
it starts from 0, 
increases to ${\sum_f{\Delta D^{B,f}(x_F,H_i|s)}}/{\sum_f{D^{B,f}(x_F,H_i|s)}}$, 
and finally tends to ${\Delta D^A(x_F,H_i|s)}/{D^A(x_F,H_i|s)}$ at $x_F\to 1$. 
We will come to this point in next section for particular hyperon 
in the specified reaction. 
We now first discuss the calculations of all these $D$'s, 
$\Delta D$'s and $N_0$ in the following. 

\subsection{Calculations of $D$'s and $N_0$}
\label{subsec:D and N0}
The contributions of hyperons 
from the different groups discussed above 
are entirely determined by the hadronization mechanisms 
in unpolarized case. 
They are independent of the polarization of the hadrons. 
We can calculate them using a hadronization model 
that gives a good description of the unpolarized data. 
For this purpose, 
the simple model used in Refs. \cite{BLM93,BL96,LB00rev} 
is a very practical choice. 
In this model, hyperons from groups (A) and (B) 
are described as the products 
of the following ``direct-formation'' 
or ``direct-fusion'' process. 
For (A), it is, 
$$(q_vq_v)^P+q_s^T\to H_i,$$ 
and for (B), it is, 
$$q_v^P+(q_sq_s)^T\to H_i,$$ 
where $q_s^T$ and $(q_sq_s)^T$ 
denote a sea quark or a sea diquark from the target. 
The number densities of the hyperons produced in these processes 
are determined by the number densities of the initial partons. 
They are given by \cite{BLM93,BL96,LB00rev}, 
\begin{eqnarray}
D^A(x_F,H_i|s) & = & \kappa_{H_i}^d f^P_D(x^P|q_vq_v)q^T_s(x^T), \\
D^{B,f}(x_F,H_i|s) & = & \kappa_{H_i} q^P_v(x^P)f^T_D(x^T|q_sq_s),
\end{eqnarray}
where, $x^P\approx x_F$ and $x^T\approx m_{H_i}^2/(sx_F)$, 
followed from energy-momentum conservation 
in the direct formation processes; 
$q_i(x)$ is the quark distribution function, 
where $q$ denotes the flavor of the quark 
and the subscript $i=v$ or $s$ denotes whether it is 
for valence or sea quarks; 
$f_D(x|q_iq_j)$ is the diquark distribution functions, 
where $q_iq_j$ denotes the flavor 
and whether they are valence or sea quarks, 
the superscripts $P$ or $T$ denote the name of the hadron; 
$\kappa_{H_i}^d$ and $\kappa_{H_i}$ are two constants 
which are fixed by fitting two data points 
in the large $x_F$ region. 

Since most of the decay process $H_j\to H_i+X$ 
that we consider are two body decay, 
$D^{C,H_j}(x_F,H_i|s)$ can be calculated 
from a convolution of $D^A(x_F,H_j|s)$ 
or $D^{B,f}(x_F,H_j|s)$ 
with the distribution describing the decay process. 
The calculations are in principle straightforward, 
but in practice a little bit complicated 
and detailed information 
of the transverse momentum distribution of $H_j$ 
is needed. 
Since the influence is not very large, 
we, for simplicity, use the following approximation. 
We neglect the distribution caused by the decay process 
and take the average value for $x_F$ instead. 
More precisely, 
we take $H_j\to H_i+M$ as an example (where $M$ denotes a meson). 
For a $H_j$ with a given longitudinal momentum fraction $x_F^j$, 
the resulting $x_F$ of the produced $H_i$ 
can take different values.
The distribution of $x_F$ at a fixed $x_F^j$ 
can be obtained from the isotropic distribution 
of the momenta of the decay products in the rest frame of $H_j$. 
This can be calculated if the transverse momentum of $H_j$ is also given.
The average value of the resulting $x_F$ 
has a simple expression, 
$\langle x_F(H_j\to H_i) \rangle=
x_F^jE_{H_i,H_j}/m_{H_j}$, 
where $E_{H_i,H_j}=(m^2_{H_j}+m^2_{H_i}-m^2_M)/(2m_{H_j})$ 
is the energy of $H_i$ in the rest frame of $H_j$. 
We see that $\langle x_F(H_j\to H_i) \rangle $ 
is independent of the transverse momentum of $H_j$. 
In our calculations, 
we simply neglect the distribution 
and take $x_F=\langle x_F(H_j\to H_i) \rangle$ 
for a given $x_F^j$. 
In this approximation, we have,
\begin{equation}
D^{C,H_j}(x_F,H_i|s)\approx Br(H_j\to H_i) \frac{m_{H_j}}{E_{H_i,H_j}} 
[D^A(\frac{m_{H_j}x_F}{E_{H_i,H_j}},H_j|s)+
\sum_fD^{B,f}(\frac{m_{H_j}x_F}{E_{H_i,H_j}},H_j|s)],
\end{equation}
where $Br(H_j\to H_i)$ is the branch ratio 
for the decay channel. 

Having calculated all these $D$'s, 
we can obtain the $N_0$ 
by parameterizing the difference of the experimental data 
on the number density $N(x_F,H_i|s)$ of produced $H_i$ 
and these $D$'s. 
We emphasized that the direct fusion model has been proposed 
to describe the production of hadrons in the fragmentation region. 
As has been shown by comparing different parts of the contributions 
to $N$ in Refs. \cite{BLM93,BL96,LB00rev}, 
this mechanism plays the dominating role in large $x_F$ region 
such as $x_F\gtrsim 0.5\sim 0.6$. 
It is clear that nobody knows \textit{a priori} 
whether the quark distribution functions can be used 
in describing the quark-fusion process 
which leads to the hadrons in the fragmentation region 
with moderately large $p_T$. 
The applicability follows from the empirical facts pointed out 
by Ochs \cite{Ochs77,Kit81}, 
and the phenomenological works by Das and Hwa \cite{Das77} 
long time ago. 
It has been pointed out \cite{Ochs77,Kit81} 
that various experiments 
have shown that the longitudinal momentum distributions 
of the produced hadrons in the fragmentation region 
are very much similar to those of the corresponding valence quarks 
in the colliding hadrons. 
The model follows directly from this observation. 
It is interesting to note 
that this simple model not only is consistent 
with the observation \cite{Ochs77,Kit81} of Ochs already in 1977 
and the theoretical works \cite{Das77} by Das and Hwa, 
but also has experienced a number of tests 
such as isospin invariance of $N_0$ etc 
(for a summary, see Ref. \cite{LB00rev}). 
Furthermore, the energy dependence that is contained 
in $D$ due to the energy dependence of $x^T$ 
leads \cite{DLL04} naturally the energy dependence 
of the single-spin left-right asymmetry $A_N$ 
observed \cite{E925} by the BNL E925 Collaboration 
compared with those by the Fermilab E704 Collaboration. 
We use this model to calculate the $D$'s and $N_0$ 
in Eq. (\ref{eq:N}). 
The quark distributions are taken from the parametrization 
at low $Q^2$ such as $Q\approx 1$ GeV/$c$.

\subsection{Calculations of $\Delta D$'s}
\label{subsec:DeltaD}
The calculations of the differences, 
i.e., $\Delta D^A$, $\Delta D^{B,f}$ 
and $\Delta D^{C,H_j}$, 
are the core ingredients of this model. 
They are described respectively in the following.

{\it Calculation of $\Delta D^A$}:
This is to determine the polarization of $H_i$ 
coming from group (A), 
i.e., $(q_vq_v)^P+q_s^T\to H_i$. 
Here, as mentioned in last section, 
we consider only the case that 
$(q_vq_v)^P+q_s^T\to H_i$ 
is associated with the production 
of $q_v+\bar q_s \to M$ where $M$ is a pseudo-scalar meson. 
We recall that according to the third point 
discussed in Section \ref{sec:picture}, 
$q_s$ should be polarized in the $-\vec{n}$ direction 
and the polarization is $-C$. 
Hence, to determine the polarization of $H_i$, 
we need to know the relative weights 
for $(q_vq_v)^P$ to be in the different 
spin states $|(q_vq_v)_{s_d,s_{dn}}\rangle$ 
where the subscripts $s_d$ and $s_{dn}$ denote 
the spin and its $n$-component of the diquark $q_vq_v$. 
Since the $q_vq_v$  is from the projectile $P$, 
these relative weights can be calculated 
using the SU(6) wave function of $P$. 
In this way, we obtain the relative weights 
for the production of $(q_vq_v)q_s$ 
in different spin states $|(q_vq_v)_{s_d,s_{dn}}q_s^\downarrow\rangle$. 
We denote this relative weight by $w(s_d,s_{dn}|q_vq_v)$.
After that, we make the projections of these different spin states 
$|(q_vq_v)_{s_d,s_{dn}}q_s^\downarrow\rangle$ 
to the wave functions $|H_i(s_n)\rangle$ of $H_i$ with 
different values of $s_n$ 
(which denotes the projection of the 
spin of $H_i$ along the $\vec n$ direction) 
and obtain the relative weights  
for $H_i$ to be in different spin states. 
The polarization of such $H_i$ is then given by, 
\begin{equation}
P^A_{H_i}=\frac{\sum_{s_n,s_d,s_{dn}} w(s_d,s_{dn}|q_vq_v)\cdot 
|\langle (q_vq_v)_{s_d,s_{dn}}q_s^\downarrow|H_i(s_n)\rangle|^2 \cdot s_n}
{\sum_{s_n,s_d,s_{dn}} w(s_d,s_{dn}|q_vq_v)\cdot 
|\langle (q_vq_v)_{s_d,s_{dn}}q_s^\downarrow|H_i(s_n)\rangle|^2 \cdot s_{n,max}}C
\equiv \alpha^A_{H_i}C.
\end{equation}
Hence, the difference, $\Delta D^A$, 
is given by,
\begin{equation}
\Delta D^A(x_F,H_i|s)=\alpha^A_{H_i}CD^A(x_F,H_i|s).
\end{equation}
For different reactions, 
the corresponding $\alpha^A_{H_i}$'s are calculated 
and the results are presented in next section.

{\it Calculation of $\Delta D^{B,f}$}:
The polarization of $H_i$ coming from group (B), 
i.e., $q_v^P+(q_sq_s)^T\to H_i$, 
is determined in the following way. 

\begin{table}
\caption{\label{tab:1}
Fragmentation spin transfer factor $t_{H_i,f}^F$ obtained 
using SU(6) wave function.}
\begin{tabular}{ccccccc}
\hline
   & $\Lambda$ &$\Sigma^+$ &$\Sigma^0$&$\Sigma^-$ &$\Xi^0$ &$\Xi^-$ \\ \hline
$u$&0&2/3&2/3&0&-1/3&0\\
$d$&0&0&2/3&2/3&0&-1/3\\
$s$&1&-1/3&-1/3&-1/3&2/3&2/3\\ \hline
\end{tabular}
\end{table}

Firstly, using the first point discussed in Section \ref{sec:picture}, 
we determine the polarization of $q_v$.
As given by Eq. (\ref{eq:P_q}), 
it is polarized in $\vec{n}$ direction 
and the polarization is $C$. 
Secondly, the corresponding relative probabilities 
to all possible spin states of ($q_sq_s$) are taken as the same. 
This means that the $q_v(q_sq_s)$ 
has the equal probability of $1/4$ 
to be in the different spin states $|q_v^\uparrow (q_sq_s)_{s_d,s_{dn}}\rangle$ 
where $s_d=0, 1$ and $s_{dn}$ takes all the possible values.
Finally, we project the different spin states of $q_v(q_sq_s)$ 
to $|H_i(s_n)\rangle$ to calculate the relative weights 
for $H_i$ to be in the different spin states.
Then the polarization of such $H_i$ is given by, 
\begin{equation}
P^{B,f}_{H_i}=\frac{\sum_{s_n,s_d,s_{dn}}
|\langle q_v^\uparrow (q_sq_s)_{s_d,s_{dn}}|H_i(s_n)\rangle|^2 \cdot s_n}
{\sum_{s_n,s_d,s_{dn}}
|\langle q_v^\uparrow (q_sq_s)_{s_d,s_{dn}}|H_i(s_n)\rangle|^2 \cdot s_{n,max}} C
\equiv \alpha^{B,f}_{H_i}C.
\end{equation}
We note that, for spin $1/2$ hyperon $H_i$, we have, 
\begin{equation}
\alpha^{B,f}_{H_i}=2\langle s_{f,n}\rangle,
\end{equation}
which is nothing else but the polarization of 
the valence quark $q$ of flavor $f$ 
in $H_i^\uparrow$ 
along the polarization of $H_i$.
It is just the fragmentation spin transfer factor $t_{H_i,f}^F$, 
which is defined \cite{LL00} as the probability for the polarization of $q_v$ to 
be transferred to $H_i$ in the fragmentation $q_v\to H_i+X$ 
in the case that the $q_v$ is contained in $H_i$.
Obviously, using the SU(6) wave function of $H_i$, 
We can calculate these $t_{H_i,f}^F$'s for different $H_i$ and quark flavor $f$. 
The results are given in Table \ref{tab:1}. 
Hence, the difference $\Delta D^{B,f}$ 
is given by,
\begin{equation}
\Delta D^{B,f}(x_F,H_i|s)=C t^F_{H_i,f}D^{B,f}(x_F,H_i|s).
\end{equation}

{\it Calculation of $\Delta D^{C,H_j}$}:
To determine the polarization of $H_i$ from the decay process 
$H_j\to H_i+X$, we should first determine the polarization 
of $H_j$. Since $H_j$ is directly produced and 
the origins belong to 
group (A) or (B) discussed above, 
its polarization is determined 
in the same way as in last two cases. 
The polarization of $H_j$ can be transferred to $H_i$ in 
the decay process $H_j\to H_i+X$. 
The probability is denoted by $t^D_{i,j}$ and 
is called the decay spin transfer factor. 
This means that, 
\begin{equation}
\Delta D^{C,H_j}(x_F,H_i|s)\approx
C t^D_{i,j} Br(H_j\to H_i) \frac{m_{H_j}}{E_{H_i,H_j}} 
[\alpha^A_{H_j}D^A(\frac{m_{H_j}x_F}{E_{H_i,H_j}},H_j|s)+
t_{H_j,f}^FD^{B,f}(\frac{m_{H_j}x_F}{E_{H_i,H_j}},H_j|s)],
\end{equation}
The decay spin transfer factor 
$t^D_{i,j}$ is universal in the sense that 
it is determined by the decay process and 
is independent of 
the process that $H_j$ is produced. 
For $\Sigma^0\to \Lambda \gamma$, 
it is determined \cite{Gatto58} as $t^D_{\Lambda,\Sigma^0}=-1/3$;
and for $\Xi\to\Lambda\pi$, 
$t^D_{\Lambda,\Xi}=(1+2\gamma)/3$, where $\gamma=0.87$ is a 
constant and can be found in Review of Particle Properties \cite{pdg}. 
 
\section{results and discussions}
\label{sec:results}

In this section, we apply the calculation method given in last section to hyperons 
in different reactions, and present the results obtained.

\subsection{Hyperon polarization in $p+p\to H_i+X$}

We first consider $pp$ or $pA$ collisions and present the results for different hyperons.
As we can see from the last section, hyperon polarizations in the proposed picture depend 
on the valence quarks of the projectile and the sea of the target. 
Hence, if we neglect the small influence from the differences 
between the sea of the proton and that of the neutron, 
we should obtain the same results for $pp$, $pn$ or $pA$ collisions.

\subsubsection{$p+p\to\Lambda +X$}
The calculations of $P_\Lambda$ in $pp\to\Lambda X$ 
have been given in Ref. \cite{LB97}. 
For completeness, we summarize the results here.

For $\Lambda$ production in $pp$ collisions, there is 
one contributing process from group (A), 
i.e., $(u_vd_v)^P+s_s^T\to\Lambda$.
There are two contributing processes from (B), i.e., 
$u_v^P+(d_ss_s)^T\to\Lambda$ and $d_v^P+(u_ss_s)^T\to\Lambda$.
Thus, we have, 
\begin{equation}
D^A(x_F,\Lambda|s) = \kappa_\Lambda^d f_D(x^P|u_vd_v)s_s(x^T), 
\end{equation}
\begin{equation}
D^{B,u}(x_F,\Lambda|s) = \kappa_\Lambda u_v(x^P)f_D(x^T|d_ss_s), 
\end{equation}
\begin{equation}
D^{B,d}(x_F,\Lambda|s) = \kappa_\Lambda d_v(x^P)f_D(x^T|u_ss_s), 
\end{equation}
where $x^P\approx x_F$ and $x^T\approx m_\Lambda^2/(sx_F)$, 
followed from energy-momentum conservation 
in the direct formation processes; 
$\kappa_\Lambda^d$ and $\kappa_\Lambda$ are two constants. 
Here, as usual, we omit the superscripts of the distribution functions 
when they are for proton. 

We take the contributions of $J^P=(1/2)^+$ hyperon decay 
into account. For $\Lambda$, we have $\Sigma^0\to\Lambda\gamma$ and 
$\Xi^{0,-}\to\Lambda\pi^{0,-}$.
For $pp\to\Sigma^0 X$, we have similar contributing processes 
from group (A) and (B) as those for $pp\to\Lambda X$, i.e., 
$(u_vd_v)^P+s_s^T\to\Sigma^0$, 
$u_v^P+(d_ss_s)^T\to\Sigma^0$ and $d_v^P+(u_ss_s)^T\to\Sigma^0$.
The number densities of $\Sigma^0$ from these processes are given by, 
\begin{equation}
D^A(x_F,\Sigma^0|s) = \kappa_{\Sigma^0}^d f_D(x^P|u_vd_v)s_s(x^T), 
\end{equation}
\begin{equation}
D^{B,u}(x_F,\Sigma^0|s) = \kappa_\Sigma u_v(x^P)f_D(x^T|d_ss_s), 
\end{equation}
\begin{equation}
D^{B,d}(x_F,\Sigma^0|s) = \kappa_\Sigma d_v(x^P)f_D(x^T|u_ss_s), 
\end{equation}
respectively. Here, $x^P\approx x_F$ and $x^T\approx m_\Sigma^2/(sx_F)$, 
$\kappa_{\Sigma^0}^d$ and $\kappa_\Sigma$ are two corresponding 
constants for $\Sigma^0$ production.
For $pp\to\Xi^0X$ and $pp\to\Xi^-X$, 
we have no contributing process from (A) 
but from (B).  They are, $u_v^P+(s_ss_s)^T\to\Xi^0$ and 
$d_v^P+(s_ss_s)^T\to\Xi^-$, respectively. 
The corresponding number densities are given by, 
\begin{equation}
D^{B,u}(x_F,\Xi^0|s) = \kappa_\Xi u_v(x^P)f_D(x^T|s_ss_s), 
\end{equation}
\begin{equation}
D^{B,d}(x_F,\Xi^-|s) = \kappa_\Xi d_v(x^P)f_D(x^T|s_ss_s), 
\end{equation}
where $x^P\approx x_F$ and $x^T\approx m_\Xi^2/(sx_F)$. 

As discussed in Section III, their contributions to $\Lambda$ 
production are given by, 
\begin{equation}
D^{C,\Sigma^0}(x_F,\Lambda|s)\approx \frac{m_{\Sigma}}{E_{\Lambda,\Sigma^0}}
[D^A(\frac{m_{\Sigma}x_F}{E_{\Lambda,\Sigma^0}},\Sigma^0|s)
+\sum_{f=u,d}{D^{B,f}(\frac{m_{\Sigma}x_F}{E_{\Lambda,\Sigma^0}},\Sigma^0|s)}],
\end{equation}
\begin{equation}
D^{C,\Xi^0}(x_F,\Lambda|s)\approx \frac{m_{\Xi}}{E_{\Lambda,\Xi}}
D^{B,u}(\frac{m_{\Xi}x_F}{E_{\Lambda,\Xi}},\Xi^0|s), 
\end{equation}
\begin{equation}
D^{C,\Xi^-}(x_F,\Lambda|s)\approx \frac{m_{\Xi}}{E_{\Lambda,\Xi}}
D^{B,d}(\frac{m_{\Xi}x_F}{E_{\Lambda,\Xi}},\Xi^-|s), 
\end{equation}
respectively. Hence, we obtain,
\begin{equation}
N(x_F,\Lambda|s)=N_0(x_F,\Lambda|s)+D^A(x_F,\Lambda|s)
+\sum_{f=u,d}D^{B,f}(x_F,\Lambda|s)
+\sum_{H_j=\Sigma^0,\Xi^0,\Xi^-}D^{C,H_j}(x_F,\Lambda|s).
\end{equation}

\begin{table}
\caption{\label{tab:2}
Relative weights $w(s_d,s_{dn}|u_vd_v)$ for 
the $u_vd_v$-diquark from proton in different spin states, 
those for the produced $\Lambda$ and $\Sigma^0$, 
the corresponding $\alpha^A_\Lambda$ and $\alpha^A_{\Sigma^0}$ 
and the total weights $w^A_\Lambda$ and $w^A_{\Sigma^0}$.}
\begin{tabular}{c|cc|cc|cc}
\hline 
Possible spin states
 & 
 \multicolumn{2}{c|}{$(u_vd_v)_{0,0}s_s^\downarrow$} & 
 \multicolumn{2}{c|}{$(u_vd_v)_{1,0}s_s^\downarrow$} & 
 \multicolumn{2}{c}{$(u_vd_v)_{1,1}s_s^\downarrow$} \\ \hline 
\begin{minipage}{4.2cm}
$w(s_d,s_{dn}|u_vd_v)$ 
\end{minipage}
 &
 \multicolumn{2}{c|}{$3/4$} & 
 \multicolumn{2}{c|}{$1/12$} & 
 \multicolumn{2}{c}{$1/6$} \\ \hline 
Possible products 
 & $\Lambda ^\downarrow $  & 
   $\Sigma^{0\downarrow }$ & 
   $\Lambda ^\downarrow $  & 
   $\Sigma^{0\downarrow }$ & 
   $\Lambda ^\uparrow $    & 
   $\Sigma^{0\uparrow }$ \\ \hline
$|\langle (q_vq_v)_{s_d,s_{dn}}q_s^\downarrow|H_i(s_n)\rangle|^2$
&  1  &  0 &  0& $1/3$& 0& $2/3$ \\ \hline
The final relative weights
 & 3/4  & 0 &  0& $1/36$& 0& $1/9$ \\ \hline
The resulted $w_{Hi}^A$ and $\alpha^A_{Hi}$ & 
 \multicolumn{3}{c}{$\Lambda$: $3/4$,  \ $-1$;} & 
 \multicolumn{3}{c}{$\Sigma^0$: $5/36$, \ $3/5$} \\ \hline 
\end{tabular}
\end{table}

The different weights $w(s_d,s_{dn}|u_vd_v)$ 
for $(u_vd_v)$ from a proton in different 
spin states can be calculated by rewriting 
the SU(6) wave function of proton as follows,
\begin{equation}
|p^{\uparrow} \rangle  =  \frac{1}{2\sqrt{3}} 
   [3u^\uparrow (ud)_{0,0}+u^\uparrow (ud)_{1,0}-
   \sqrt{2}u^\downarrow (ud)_{1,1}]. 
\end{equation}
Taking into account that the proton is unpolarized thus 
has equal probabilities to be in $|p^\uparrow\rangle$ and 
$|p^\downarrow\rangle$, we obtain the results 
for $w(s_d,s_{dn}|u_vd_v)$ 
as shown in Table \ref{tab:2}.
We then project the different spin states of 
$(u_vd_v)s_s$ on the wave function of $\Lambda$ 
and obtain the relative weights for the production of 
$\Lambda$ from this process in different spin states in Table \ref{tab:2}. 
From these results, we obtain also the $\alpha_\Lambda^A$ as shown in the table. 
Hence its contribution to $\Delta N(x_F,\Lambda|s)$, i.e., 
$\Delta D^A(x_F,\Lambda|s)=C \alpha_\Lambda^A D^A(x_F,\Lambda|s)$ 
can also be calculated. 
Similarly, we obtain the corresponding results for $\Sigma^0$ production
and show them in the same table. From these results, we obtain that, 
\begin{equation}
\Delta N(x_F,\Lambda|s)=C \alpha^A_\Lambda D^A(x_F,\Lambda|s)
+\sum_{H_j=\Sigma^0,\Xi^0,\Xi^-}\Delta D^{C,H_j}(x_F,\Lambda|s),
\end{equation}
\begin{equation}
\Delta D^{C,\Sigma^0}(x_F,\Lambda|s)\approx 
C t^D_{\Lambda,\Sigma^0} \frac{m_{\Sigma}}{E_{\Lambda,\Sigma^0}}
[\alpha^A_{\Sigma^0} D^A(\frac{m_{\Sigma}x_F}{E_{\Lambda,\Sigma^0}},\Sigma^0|s)
+\sum_{f=u,d}{t^F_{\Sigma^0,f} D^{B,f}(\frac{m_{\Sigma}x_F}{E_{\Lambda,\Sigma^0}},
\Sigma^0|s)}],
\end{equation}
\begin{equation}
\Delta D^{C,\Xi^0}(x_F,\Lambda|s)\approx
C t^D_{\Lambda,\Xi} t^F_{\Xi^0,u} \frac{m_{\Xi}}{E_{\Lambda,\Xi}}
D^{B,u}(\frac{m_{\Xi}x_F}{E_{\Lambda,\Xi}},\Xi^0|s),
\end{equation}
\begin{equation}
\Delta D^{C,\Xi^-}(x_F,\Lambda|s)\approx
C t^D_{\Lambda,\Xi} t^F_{\Xi^-,d} \frac{m_{\Xi}}{E_{\Lambda,\Xi}}
D^{B,d}(\frac{m_{\Xi}x_F}{E_{\Lambda,\Xi}},\Xi^-|s),
\end{equation}
where the fragmentation spin transfer factors $t^F_{H_j,f}$'s 
for different quark flavor $f$ and different hyperon $H_j$ are 
listed in Table \ref{tab:1}, and the decay spin transfer factors 
$t^D_{\Lambda,H_j}$'s are given in Section III.

Using the results given by Eqs. (15--31), we can now calculate 
$P_\Lambda(x_F|s)$ as a function of $x_F$. 
Before we show the numerical results, 
we first look at the qualitative features. 
Just as mentioned at the end of Subsection \ref{subsec:general}, 
for $pp\to\Lambda X$, we expect that 
$D^A(x_F,\Lambda|s)$, i.e., the contribution from $(u_vd_v)^P+s_s^T\to\Lambda$, 
dominates $N(x_F,\Lambda|s)$ at $x_F\to 1$. 
$N_0(x_F,\Lambda|s)$ dominates for very small $x_F$, 
while $D^{B,u}(x_F,\Lambda|s)$ and $D^{B,d}(x_F,\Lambda|s)$, 
i.e., those from $u_v^P+(d_ss_s)^T\to\Lambda$ and $d_v^P+(u_ss_s)^T\to\Lambda$, 
play the dominating role for moderate $x_F$. 
Since $\alpha^A_\Lambda=-1$ and $t^F_{\Lambda,u}=t^F_{\Lambda,d}=0 $, 
we expect that, 
if the decay contribution can be neglected, 
for $x_F$ going from 0 to 1, 
$P_\Lambda(x_F|s)$ starts from 0, becomes nonzero quite slowly 
and tends to $C\alpha^A_\Lambda=-C$ at $x_F=1$. 
Taking the decay contribution into account, 
we expect a significant contribution from $\Sigma^0\to\Lambda\gamma$ 
which leads to a negative $P_\Lambda(x_F|s)$ at moderate $x_F$ 
and makes the $|P_\Lambda(x_F|s)|$ less than $C$ at $x_F$ near 1. 

We now use Eqs. (15--31) to get the numerical results 
for $P_\Lambda(x_F|s)$ as a function of $x_F$. 
The only unknown parameters are the $\kappa$'s 
which should be determined by the unpolarized 
experimental data on the number densities 
for the corresponding hyperons. 
To reduce the arbitrariness in determining these $\kappa$'s, 
we take the same $\kappa_{H_i}$'s for different 
hyperons produced in processes of group (B). 
For those from group (A), we take the 
$\kappa_{H_i}^d$'s for different hyperons as, 
\begin{equation}
\kappa_{H_i}^d=w^d_{H_i}\cdot \kappa^d,
\end{equation}
where $\kappa^d$ is taken as a constant independent of $H_i$, and
\begin{equation}
w^d_{H_i}\equiv \sum_{s_d,s_{dn},s_n}
w(s_d,s_{dn}|q_vq_v)\cdot 
|\langle (q_vq_v)_{s_d,s_{dn}}q_s^\downarrow|H_i(s_n)\rangle|^2,
\end{equation}
is the total relative weight for the production of $H_i$ 
from $(q_vq_v)^P+q_s^T\to H_i$.
This means that 
only the $H_i$-dependence of $\kappa^d_{H_i}$ 
from the spin statistics is taken into account.
The $w_{H_i}^d$'s for $\Lambda$ and $\Sigma^0$ are given in Table \ref{tab:2}.
In this way, we have only two free $\kappa$'s, i.e., $\kappa^d$ 
for $(q_vq_v)^P+q_s^T\to H_i$ 
and $\kappa_{Hi}\equiv\kappa$ 
for $q_v^P+(q_sq_s)^T\to H_i$. 
We determine \cite{ft:kappa} 
them by using the data for $N(x_F,\Lambda|s)$. 

\begin{figure}
\includegraphics[]{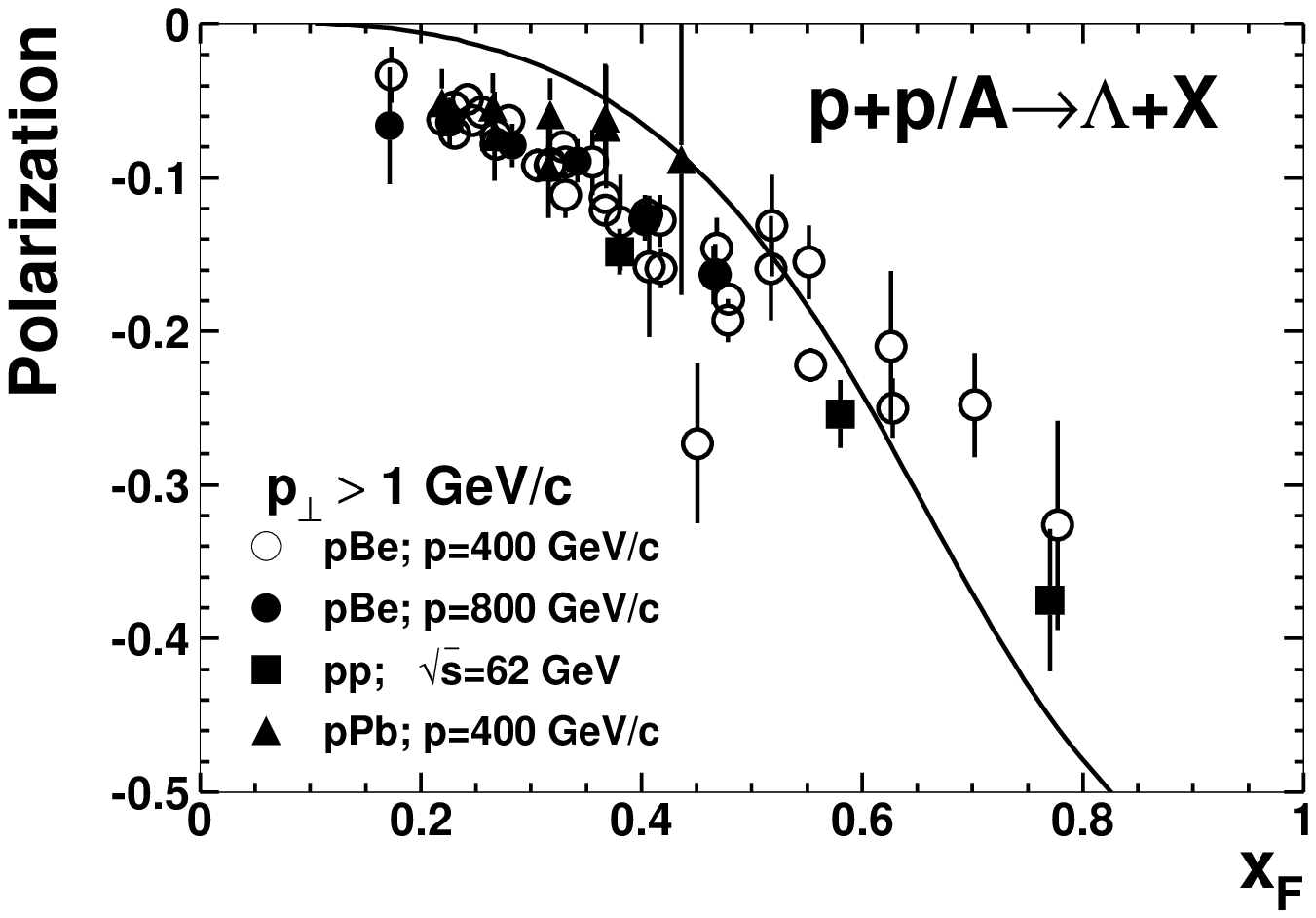}
\caption{\label{fig:pp2Lambda}
Calculated results for the polarization of $\Lambda$ in $pp$ collisions 
as a function of $x_F$. 
Data are taken from Refs. \cite{Smi87,Lun89,Ram94}. }
\end{figure}

Having the $\kappa$'s, we calculate $P_\Lambda(x_F|s)$ as a function of $x_F$.
For the unpolarized quark distribution functions,
we use the GRV 98 LO set \cite{GRV98}. 
For the unpolarized diquark distribution $f_D(x^P|q_vq_v)$, 
we use the parametrization given in Ref. \cite{Dug93}.
For the unpolarized diquark distribution $f_D(x^T|q_sq_s)$, 
we simply use a convolution of $q_s(x)$ for the two sea quarks.
The obtained results for $P_\Lambda(x_F|s)$ are given in Fig. \ref{fig:pp2Lambda}.
The data in the figure are only those for $p_T>1$GeV/$c$, 
because the $C$ that we used in the calculations are determined 
by $A_N$ for the $p_T$ interval $0.7<p_T<2.0$ GeV/$c$. 
We see clearly that, as $x_F$ changes from 0 to 1, 
the obtained $P_\Lambda(x_F|s)$ indeed starts from 0, 
goes slowly to about $-15\%$ 
and finally to about $-50\%$ at $x_F\to 1$. 
These qualitative features 
are the same as we expect from the above-mentioned qualitative analysis 
and are in good agreement with the data \cite{Smi87,Lun89,Ram94}.

It should be mentioned that, in the calculations presented above, 
we considered only the associated production of hyperons of group (A), 
i.e., $(q_vq_v)^P+q_s^T\to H_i$, with a pseudo-scalar meson $M$, 
i.e., $q_v^P+\bar{q}_s^T\to M$. 
It is clear that the associatively produced hadron can also be a vector meson. 
Taking this into account, we expect that the correlations 
between the spins of these quarks will be reduced. 
As a consequence, the polarizations of hyperons from group (A) 
should be slightly reduced. 
Since such mesons contribute mainly at large $x_F$, 
this effect should make $|P_\Lambda|$ smaller than those presented 
in Fig. \ref{fig:pp2Lambda} in the large $x_F$ region. 
From the figure, we also see that there is indeed room left 
for such an effect. 
Similar effects exist for 
$pp\to\Sigma^+X$ and $\Sigma^-p\to\Sigma^-(\mbox{or}\ \Xi^-)X$ 
that will be discussed in the following. 
How large this effect can be is determined by the hadronization mechanisms. 
Presently, we are working on an estimation. 
The results will be published separately \cite{DLfuture}.

\subsubsection{$p+p\to\Sigma^{\pm}$(or $\Xi^{0,-}$)$+X$}
In a completely similar way, we calculate hyperon polarizations for 
$pp\to\Sigma^{\pm}\mbox{or}\ \Xi^{0,-})X$. The results are given in the following.

\begin{table}
\caption{\label{tab:3}
Relative weights $w(s_d,s_{dn}|u_vu_v)$ for 
the $u_vu_v$-diquark from proton in different spin states, 
those for the produced $\Sigma^+$,  
the corresponding $\alpha^A_{\Sigma^+}$ and total relative weight $w^A_{\Sigma^+}$.}
\begin{tabular}{c|c|c}
\hline 
Possible spin states
 & $(u_vu_v)_{1,0}s_s^\downarrow$ & $(u_vu_v)_{1,1}s_s^\downarrow$ \\ \hline 
$w(s_d,s_{dn}|u_vu_v)$  & $1/3$ & $2/3$ \\ \hline 
Possible products  
   & $\Sigma^{+\downarrow }$ & $\Sigma^{+\uparrow }$ \\ \hline
$|\langle (u_vu_v)_{s_d,s_{dn}}s_s^\downarrow|\Sigma^+(s_n)\rangle|^2$
   &  1/3  & 2/3 \\ \hline
The final relative weights & 1/9  & 4/9   \\ \hline
The resulted $w_{\Sigma^+}^A$ and $\alpha^A_{\Sigma^+}$ 
  & \multicolumn{2}{c}{$w_{\Sigma^+}^A=5/9$,\ $\alpha^A_{\Sigma^+}=3/5$} \\ \hline 
\end{tabular}
\end{table}

For $pp\to\Sigma^+X$, there are one contributing process from group (A), i.e., 
$(u_vu_v)^P+s_s^T\to\Sigma^+$ and one from group (B), i.e., 
$u_v^P+(u_ss_s)^T\to\Sigma^+$.
The corresponding number densities are given by,
\begin{equation}
D^A(x_F,\Sigma^+|s)=\kappa^d_{\Sigma^+}f_D(x^P|u_vu_v)s_s(x^T),
\end{equation}
\begin{equation}
D^{B,u}(x_F,\Sigma^+|s)=\kappa u_v(x^P)f_D(x^T|u_ss_s),
\end{equation}
where $x^P\approx x_F$ and $x^T\approx m_\Sigma^2/(sx_F)$.
In the case that only $J^P=(1/2)^+$ hyperon decay is taken into account, 
there is no contribution from hyperon decay to $\Sigma^+$ production. 
Hence, we have,
\begin{equation}
N(x_F,\Sigma^+|s)=N_0(x_F,\Sigma^+|s)+D^A(x_F,\Sigma^+|s)+D^{B,u}(x_F,\Sigma^+|s),
\end{equation}
\begin{equation}
\Delta N(x_F,\Sigma^+|s)=
C[\alpha^A_{\Sigma^+}D^A(x_F,\Sigma^+|s)+t^F_{\Sigma^+,u}D^{B,u}(x_F,\Sigma^+|s)],
\end{equation}
where $t^F_{\Sigma^+,u}$ is given in Table \ref{tab:1}. 
$\alpha_{\Sigma^+}^A$ is calculated in completely the same way 
as that for $\alpha_{\Lambda}^A$ presented in last subsection. 
The results are given in Table \ref{tab:3}. Thus, we have, 
\begin{equation}
P_{\Sigma^+}(x_F|s)=C\frac{\alpha^A_{\Sigma^+}D^A(x_F,\Sigma^+|s)+t^F_{\Sigma^+,u}D^{B,u}(x_F,\Sigma^+|s)}
{N_0(x_F,\Sigma^+|s)+D^A(x_F,\Sigma^+|s)+D^{B,u}(x_F,\Sigma^+|s)}.
\end{equation}

The situations for the productions of $\Sigma^-$, 
$\Xi^0$ and $\Xi^-$ in $pp$ collisions 
are even simpler:
There is no contributing process from group (A) but 
only one contributing process from (B). 
When only $J^P=(1/2)^+$ hyperon decay is taken into account, 
there is also no contribution from hyperon decay to these hyperons. 
Hence, we have,
\begin{equation}
P_{\Sigma^-}(x_F|s)=C\frac{t^F_{\Sigma^-,d}D^{B,d}(x_F,\Sigma^-|s)}
{N_0(x_F,\Sigma^-|s)+D^{B,d}(x_F,\Sigma^-|s)},
\end{equation}
\begin{equation}
P_{\Xi^0}(x_F|s)=C\frac{t^F_{\Xi^0,u}D^{B,u}(x_F,\Xi^0|s)}
{N_0(x_F,\Xi^0|s)+D^{B,u}(x_F,\Xi^0|s)},
\end{equation}
\begin{equation}
P_{\Xi^-}(x_F|s)=C\frac{t^F_{\Xi^-,d}D^{B,d}(x_F,\Xi^-|s)}
{N_0(x_F,\Xi^-|s)+D^{B,d}(x_F,\Xi^-|s)},
\end{equation}
where the $t^F_{H_i,f}$'s are given in Table \ref{tab:1}, 
\begin{equation}
D^{B,d}(x_F,\Sigma^-|s)=\kappa d_v(x^P)f_D(x^T|d_ss_s),
\end{equation}
and $D^{B,u}(x_F,\Xi^0|s)$ and $D^{B,d}(x_F,\Xi^-|s)$ 
are given by Eqs. (21) and (22).

From these equations, we can calculate 
$P_{\Sigma^{\pm}}(x_F|s)$ and $P_{\Xi}(x_F|s)$ in $pp$ collisions. 
Just as we mentioned at the end of Subsection \ref{subsec:general}, 
for $pp\to\Sigma^+X$, we expect that 
$D^A(x_F,\Sigma^+|s)$ dominates at large $x_F$, 
$D^{B,u}(x_F,\Sigma^+|s)$ plays the dominating role at moderate $x_F$. 
Hence, we expect from Eq. (38) that, 
$P_{\Sigma^+}(x_F|s)=0$ for $x_F$ near 0, 
increases to $Ct^F_{\Sigma^+,u}=0.4$
and finally tends to $C\alpha^A_{\Sigma^+}=0.36$ with increasing $x_F$. 
The situations for $\Sigma^-$, $\Xi^0$ and $\Xi^-$ are even simpler 
since there is no contributing process from group (A). 
Here, $N_0$ dominates at small $x_F$, while $D^B$ dominates at large $x_F$. 
We thus expect that 
$P_{\Sigma^-}(x_F|s)=0$ for $x_F$ near 0 
and increases to $Ct^F_{\Sigma^-,d}=0.4$ with increasing $x_F$. 
For $\Xi$, $P_{\Xi}(x_F|s)=0$ for $x_F$ near 0 
and increases to $Ct^F_{\Xi,f}=-0.2$ with increasing $x_F$. 
To summarize, we expect that both $P_{\Sigma^+}(x_F|s)$ and $P_{\Sigma^-}(x_F|s)$ 
are positive in sign and increase fast to about $40\%$ with increasing $x_F$. 
On the other hand, both $P_{\Xi^0}(x_F|s)$ and $P_{\Xi^-}(x_F|s)$ 
are negative in sign, 
and they decrease to $-20\%$ with increasing $x_F$. 
The magnitudes of $P_{\Sigma^{\pm}}$ are expected to be larger than those of $P_\Xi$. 
These qualitative features are in agreement 
with the available data \cite{Wil87,Dec83,Ramei86,Hel83,Mor95,Dur91}. 

\begin{figure}
\includegraphics[]{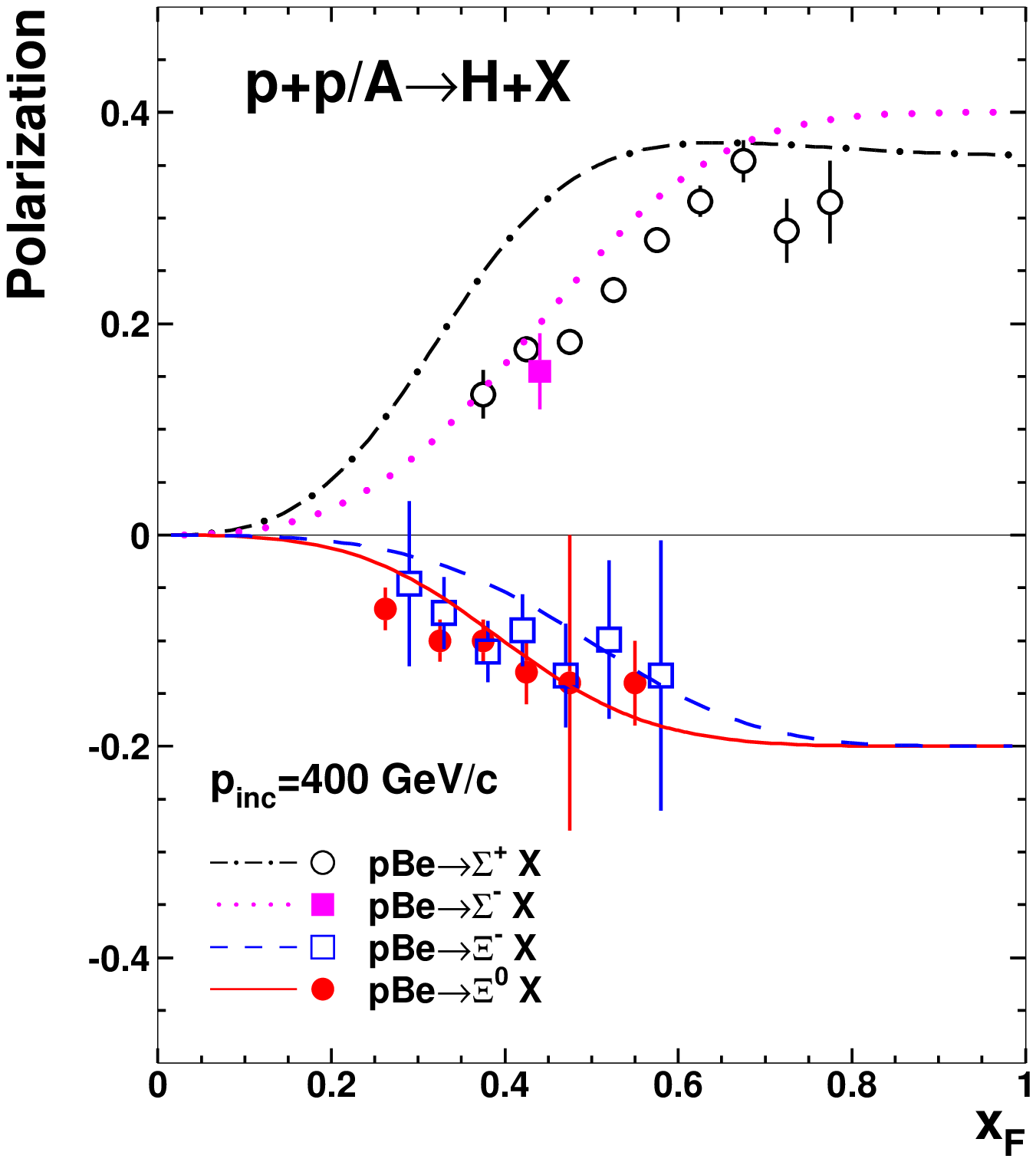}
\caption{\label{fig:pp2other400}
Calculated results for the polarizations of $\Sigma^{\pm}$ and $\Xi^{0,-}$ 
at $p_{inc}=400\ \mbox{GeV}/c$ in $pp$ collisions as functions of $x_F$. 
Data are taken from Refs. \cite{Wil87,Dec83,Ramei86,Hel83}. }
\end{figure}

\begin{figure}
\includegraphics[]{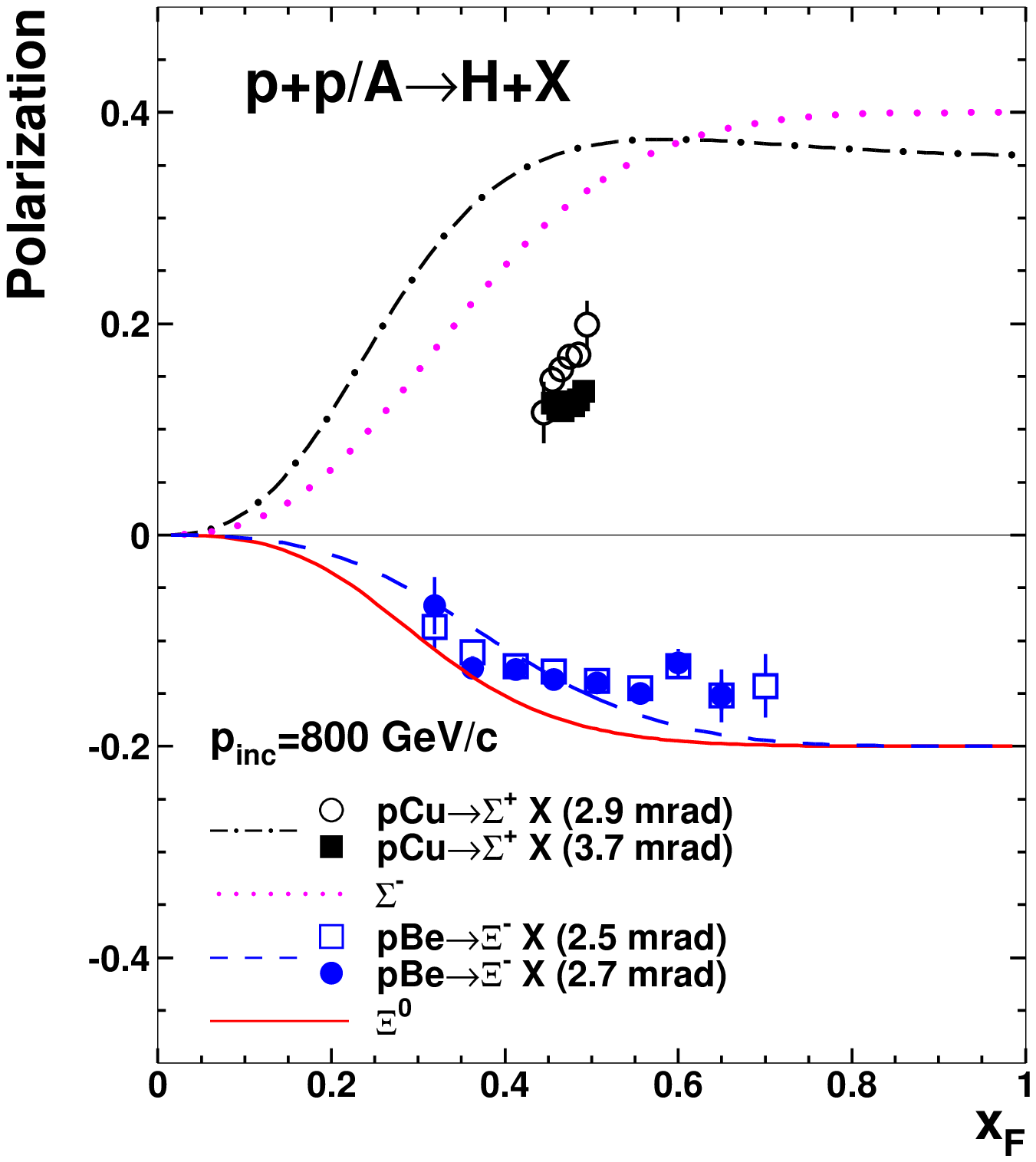}
\caption{\label{fig:pp2other800}
Calculated results for the polarizations of $\Sigma^{\pm}$ and $\Xi^{0,-}$ 
at $p_{inc}=800\ \mbox{GeV}/c$ in $pp$ collisions as functions of $x_F$. 
Data are taken from Refs. \cite{Mor95,Dur91}. }
\end{figure}

We now use Eqs. (38--41) to obtain the numerical results 
for $P_{\Sigma^{\pm}}(x_F|s)$ and $P_{\Xi}(x_F|s)$ in $pp$ collisions. 
The only unknown thing is $N_0$ for the corresponding hyperon. 
As we mentioned earlier in Section III, $N_0$ is independent of 
the polarization properties and can be determined 
using the data for $N(x_F,H_i|s)$. 
Since there is no suitable data available for different hyperons, 
we make the following estimation based on 
our result on $N_0(x_F,\Lambda|s)$.
We simply assume that the $x_F$-dependences 
of all these $N_0(x_F,H_i|s)$'s are the same. 
Hence, we have 
$N_0(x_F,H_i|s)=N_0(x_F,\Lambda|s)\langle n_{H_i}\rangle/\langle n_\Lambda\rangle$, 
where $\langle n_{H_i}\rangle$ is the average number of $H_i$ in the central region. 
For directly produced hyperons, 
we take $\langle n_{\Sigma^+}\rangle=\langle n_{\Sigma^-}\rangle
=\langle n_{\Sigma^0}\rangle=\langle n^{dir}_\Lambda\rangle$, 
and $\langle n_{\Xi}\rangle=\lambda\langle n_{\Sigma}\rangle$, 
where $\lambda=0.3$ denotes the strangeness suppression factor. 
We take $\Sigma^0$ and $\Xi$ decays into account, 
and obtain that 
$\langle n_{\Lambda}\rangle=(2+2\lambda)\langle n^{dir}_{\Lambda}\rangle$. 
In this way, we obtain a rough estimation of the $N_0(x_F,H_i|s)$. 
Using this, we obtain the numerical results for $P_{\Sigma^{\pm}(\mbox{or}\ \Xi)}(x_F|s)$ 
at $p_{inc}=400$ GeV/$c$ as shown in Fig. \ref{fig:pp2other400}, 
and those at $p_{inc}=800$ GeV/$c$ in Fig. \ref{fig:pp2other800}. 
We see 
that the results show clearly the qualitative features mentioned above 
and that these features are in agreement 
with the data \cite{Wil87,Dec83,Ramei86,Hel83,Mor95,Dur91}. 

\subsection{Hyperon polarization in $K^-+p\to H_i+X$}

There exist also data for hyperon polarization in $K^-p\to\Lambda X$. 
The data show that $P_\Lambda$ in this process 
is also significantly different from zero for large $x_F$. 
Furthermore, compared with those for $P_\Lambda$ in $pp\to\Lambda X$, 
the data show that $P_\Lambda$ in $K^-p\to\Lambda X$ 
has different sign from that for $pp\to\Lambda X$. 
For $x_F$ from 0 to 1, $P_\Lambda$ in $K^-p\to\Lambda X$ 
begins with $P_\Lambda\approx 0$, increases monotonically to 
about $40\%$ at $x_F\to 1$. 
Now, we apply the proposed picture to this process, 
compare the results with the data, 
and make predictions for other hyperons. 


\subsubsection{$K^-+p\to\Lambda +X$}
Since we have a meson as projectile, 
there is no contributing process from group (A) to $K^-p\to\Lambda X$. 
There is one contributing process from (B), i.e., 
$s_v^P+(u_sd_s)^T\to\Lambda$.
Thus, we have, 
\begin{equation}
D^{B,s}(x_F,\Lambda|s) = \kappa s^K_v(x^P)f_D(x^T|u_sd_s), 
\end{equation}
where $x^P\approx x_F$ and $x^T\approx m_\Lambda^2/(sx_F)$. 
The superscript $K$ for the quark distribution functions 
denotes that they are for quarks in $K$ meson. 

Just as $pp\to\Lambda X$, there are also contributions 
from hyperon decays to $K^-p\to\Lambda X$. 
In $K^-p$ collisions, we have similar contributing processes 
from group (B) to $\Sigma^0$ and $\Xi$ production, i.e., 
$s_v^P+(u_sd_s)^T\to\Sigma^0$, 
$s_v^P+(u_ss_s)^T\to\Xi^0$ and 
$s_v^P+(d_ss_s)^T\to\Xi^-$, respectively. 
The corresponding number densities are given by, 
\begin{equation}
D^{B,s}(x_F,\Sigma^0|s) = \kappa s^K_v(x^P)f_D(x^T|u_sd_s), 
\end{equation}
\begin{equation}
D^{B,s}(x_F,\Xi^0|s) = \kappa s^K_v(x^P)f_D(x^T|u_ss_s), 
\end{equation}
\begin{equation}
D^{B,s}(x_F,\Xi^-|s) = \kappa s^K_v(x^P)f_D(x^T|d_ss_s), 
\end{equation}
respectively. 

Hence, we obtain,
\begin{equation}
P_\Lambda(x_F|s)=C \frac
{t^F_{\Lambda,s} D^{B,s}(x_F,\Lambda|s)
+\sum_{H_j=\Sigma^0,\Xi^0,\Xi^-}t^D_{\Lambda,H_j} t^F_{H_j,s} D^{C,H_j}(x_F,\Lambda|s)}
{N_0(x_F,\Lambda|s)+D^{B,s}(x_F,\Lambda|s)
+\sum_{H_j=\Sigma^0,\Xi^0,\Xi^-}D^{C,H_j}(x_F,\Lambda|s)},
\end{equation}
\begin{equation}
D^{C,H_j}(x_F,\Lambda|s)\approx \frac{m_{H_j}}{E_{\Lambda,H_j}}
D^{B,s}(\frac{m_{H_j}x_F}{E_{\Lambda,H_j}},H_j|s),
\end{equation}
where the fragmentation spin transfer factors $t^F_{H_j,s}$'s 
for quark flavor $s$ and different hyperon $H_j$ are 
given in Table \ref{tab:1}, and the decay spin transfer factors 
$t^D_{\Lambda,H_j}$'s are given in Section III.

Using Eq. (47), 
we can now calculate $P_\Lambda(x_F|s)$ in $K^-p$ collisions 
as a function of $x_F$. 
Just as mentioned at the end of Subsection \ref{subsec:general}, 
for $K^-p\to\Lambda X$, we expect that $N_0$ dominates at small $x_F$, 
while $D^B(x_F,\Lambda|s)$ dominates at large $x_F$ 
since there is no contributing process from group (A). 
It is clear that $P_\Lambda(x_F|s)>0$ and, 
if the decay contribution is neglected, 
$P_\Lambda(x_F|s)=0$ for $x_F$ near 0 
and increases to $C t^F_{\Lambda,s}=C=0.6$ with increasing $x_F$. 
The decay contribution should make $P_\Lambda(x_F|s)$ 
slightly less than $C$ at $x_F\to 1$ 
because $0<t^D_{\Lambda,H_j} t^F_{H_j,s}<t^F_{\Lambda,s}$ 
for all the $H_j=\Sigma^0,\Xi^0,\Xi^- $. 

\begin{figure}
\includegraphics[]{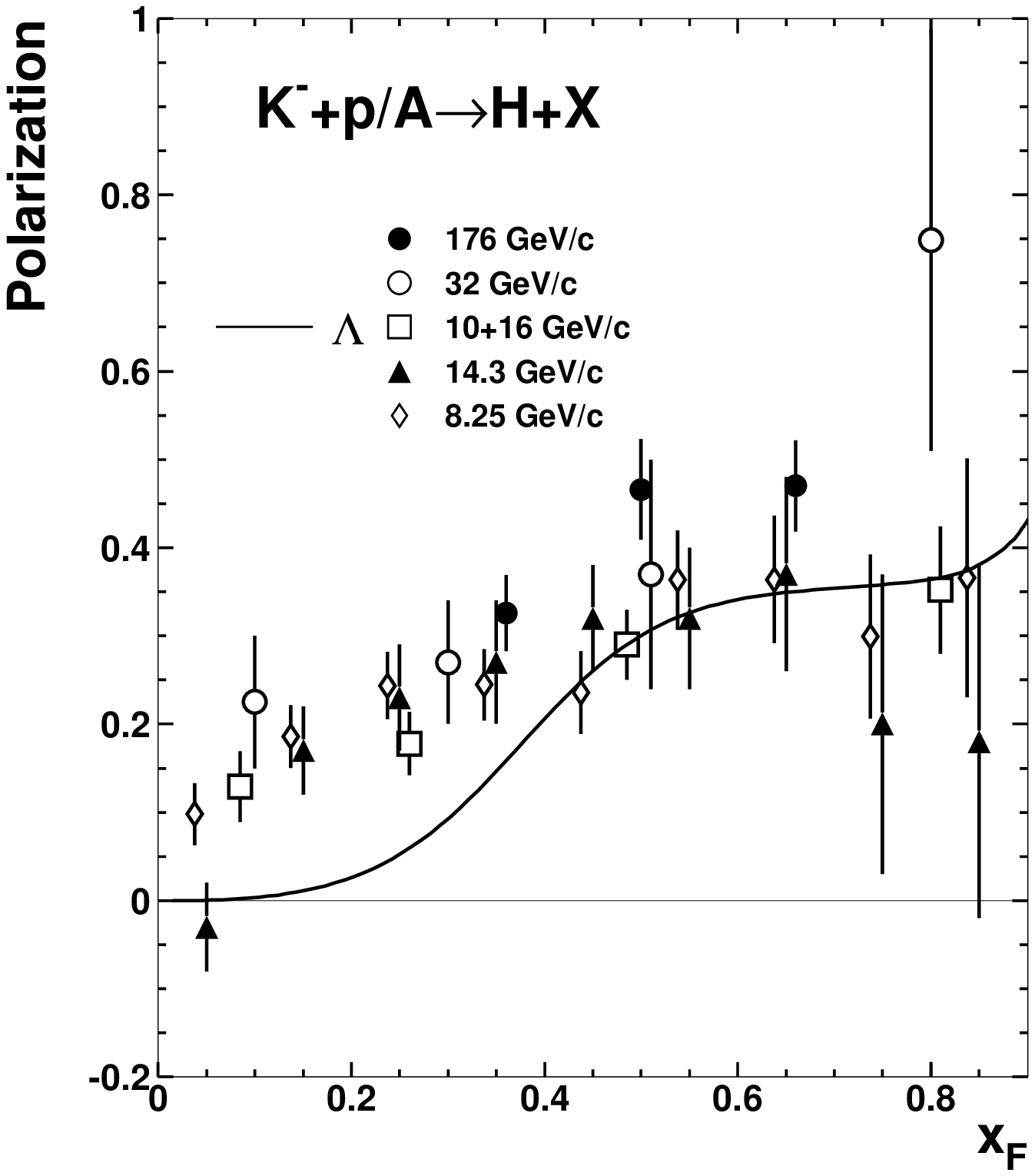}
\caption{\label{fig:K-p2Lambda}
Calculated results for the polarization of $\Lambda$ in $K^-p$ collisions 
as a function of $x_F$. 
Data are taken from Refs. \cite{Gou86,Fac79,Gra78,Abr76,Bau79}. }
\end{figure}

We now use Eq. (47) to obtain the numerical results 
for $P_\Lambda(x_F|s)$ in $K^-p$ collisions 
in order to get a more precise feeling 
of the above-mentioned qualitative features. 
Here, for the unpolarized quark distribution function of kaon, 
we use the GRV-P LO set \cite{GRV92} for that of pion instead. 
For $N_0(x_F,H_i|s)$ in this process, we simply assume it to be the same as 
that in $pp$ collisions. 
The obtained results for $P_\Lambda(x_F|s)$ are given in Fig. \ref{fig:K-p2Lambda}. 
We see that clearly as $x_F$ increases from 0 to 1, 
the obtained $P_\Lambda(x_F|s)$ starts from 0 and increases to about $40\%$ at $x_F\to 1$. 
These qualitative features are in good agreement with the data 
\cite{Gou86,Fac79,Gra78,Abr76,Bau79}. 

\subsubsection{$K^-+p\to\Sigma^{\pm}$ (or $\Xi^{0,-}$)$+X$}

Similar to $K^-p\to\Lambda X$ in $K^-p$ collisions, 
there is also only one contributing process from group (B) 
to the production of each of these hyperons. 
For $\Sigma^+$ and $\Sigma^-$, they are 
$s_v^P+(u_su_s)^T\to\Sigma^+$ and $s_v^P+(d_sd_s)^T\to\Sigma^-$, respectively. 
The corresponding number densities are given by,
\begin{equation}
D^{B,s}(x_F,\Sigma^+|s)=\kappa s^K_v(x^P)f_D(x^T|u_su_s),
\end{equation}
\begin{equation}
D^{B,s}(x_F,\Sigma^-|s)=\kappa s^K_v(x^P)f_D(x^T|d_sd_s),
\end{equation}
respectively. 
The contributing processes from (B) to $\Xi$ production 
have been given in the last subsection 
and their corresponding number densities are given by Eqs. (45) and (46). 
Thus, we have,
\begin{equation}
P_{\Sigma^+}(x_F|s)=C\frac{t^F_{\Sigma^+,s}D^{B,s}(x_F,\Sigma^+|s)}
{N_0(x_F,\Sigma^+|s)+D^{B,s}(x_F,\Sigma^+|s)},
\end{equation}
\begin{equation}
P_{\Sigma^-}(x_F|s)=C\frac{t^F_{\Sigma^-,s}D^{B,s}(x_F,\Sigma^-|s)}
{N_0(x_F,\Sigma^-|s)+D^{B,s}(x_F,\Sigma^-|s)},
\end{equation}
\begin{equation}
P_{\Xi^0}(x_F|s)=C\frac{t^F_{\Xi^0,s}D^{B,s}(x_F,\Xi^0|s)}
{N_0(x_F,\Xi^0|s)+D^{B,s}(x_F,\Xi^0|s)},
\end{equation}
\begin{equation}
P_{\Xi^-}(x_F|s)=C\frac{t^F_{\Xi^-,s}D^{B,s}(x_F,\Xi^-|s)}
{N_0(x_F,\Xi^-|s)+D^{B,s}(x_F,\Xi^-|s)},
\end{equation}
where the $t^F_{H_i,s}$'s are given in Table \ref{tab:1}. 

It is also clear that, for the productions of these hyperons, 
$N_0$ dominates at small $x_F$, 
while $D^B$ plays the dominating role at large $x_F$. 
We expect that, for $\Sigma^{\pm}$ in $K^-p$ collisions, 
$P_{\Sigma^{\pm}}(x_F|s)=0$ for $x_F$ near 0 
and decreases to $Ct^F_{\Sigma,s}=-0.2$ with increasing $x_F$. 
For $K^-p\to\Xi X$, $P_{\Xi}(x_F|s)=0$ for $x_F$ near 0 
and increases to $Ct^F_{\Xi,s}=0.4$ with increasing $x_F$. 
This means that $P_{\Sigma^{\pm}}(x_F|s)$ is negative in sign 
and decreases to $-20\%$ with increasing $x_F$, 
while $P_{\Xi}(x_F|s)$ is positive in sign 
and increases fast to $40\%$ with increasing $x_F$. 
The magnitudes of $P_{\Sigma^{\pm}}$ should be smaller 
than those of $P_{\Xi}$. 
These qualitative features are different from those for $pp$ collisions 
and can be checked by future experiments. 

\begin{figure}
\includegraphics[]{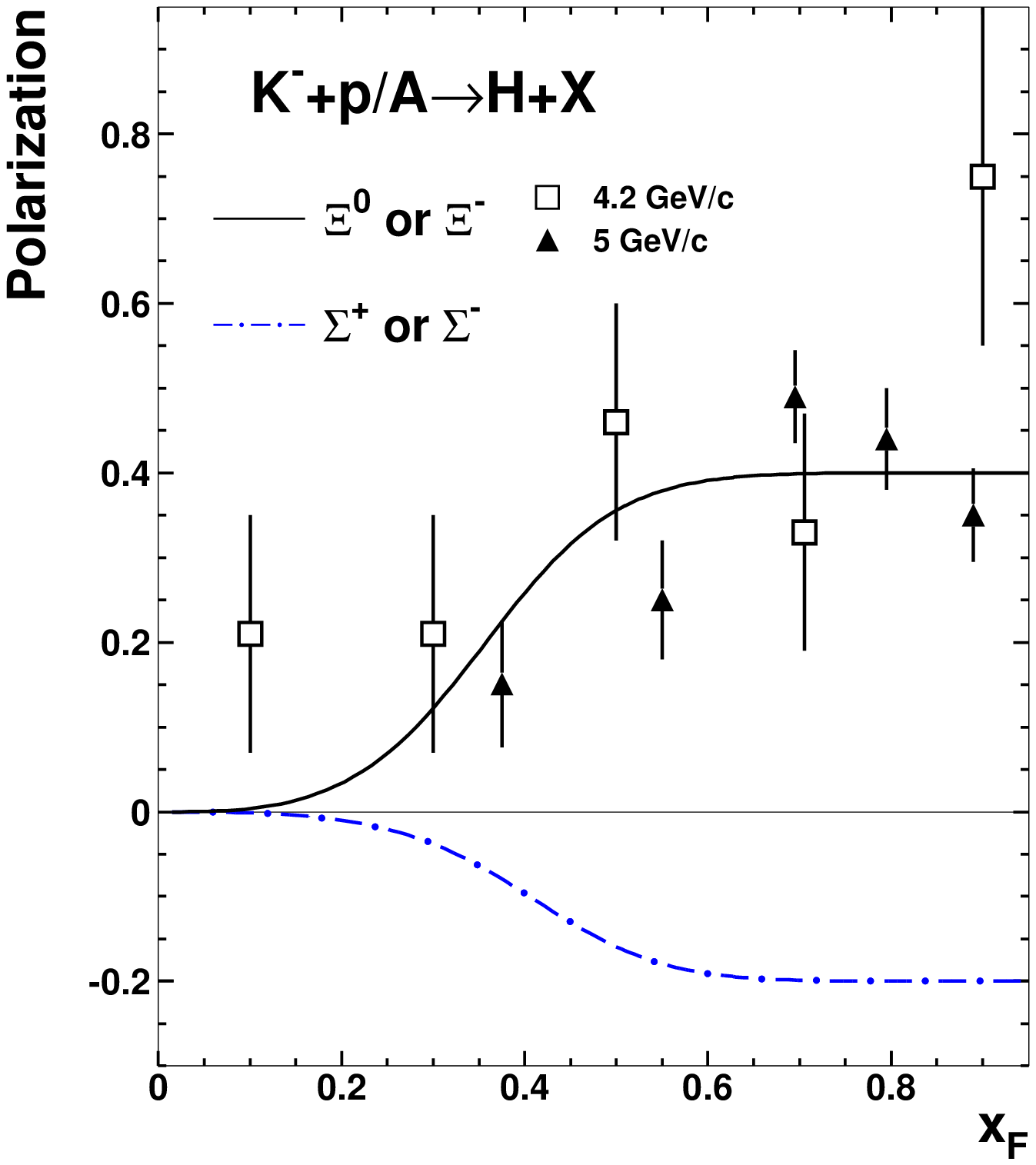}
\caption{\label{fig:K-p2other}
Calculated results for the polarizations of $\Sigma^{\pm}$ and $\Xi^{0,-}$ 
in $K^-p$ collisions as functions of $x_F$. 
Data for $\Xi^-$ are taken from Refs. \cite{Gan77,Ben85}. }
\end{figure}

By using Eqs. (51--54), we also obtain the numerical results for 
$P_{\Sigma^{\pm}}(x_F|s)$ and $P_{\Xi}(x_F|s)$ as functions of $x_F$ in $K^-p$ collisions. 
They are given in Fig. \ref{fig:K-p2other}. 
At present, there are only data available for $\Xi^-$ \cite{Gan77,Ben85} 
among these hyperons. 
We see that the results show clearly the qualitative features mentioned above 
and that those for $\Xi^-$ are consistent with the available data. 
They all can be further checked by further experiments. 

\subsection{Hyperon polarization in $\pi^\pm+p\to H_i+X$}

Now, we apply the proposed picture to $\pi^\pm p$ collisions. 
From the isospin symmetry, 
we obtain that $P_\Lambda$ in $\pi^+p\to\Lambda X$ is the same as 
that in $\pi^-p\to\Lambda X$, 
$P_{\Sigma^+}$ and $P_{\Xi^0}$ in $\pi^+p$ collisions are the same as 
$P_{\Sigma^-}$ and $P_{\Xi^-}$ in $\pi^-p$ collisions, respectively. 
So we only give the calculations for $\pi^-+p\to H_i+X$. 

\subsubsection{$\pi^-+p\to\Lambda +X$}

Similar to $K^-p\to\Lambda X$, there is also no contributing process 
from group (A) to $\pi^-p\to\Lambda X$. 
There is one contributing process from (B), i.e., 
$d_v^P+(u_ss_s)^T\to\Lambda$.
Thus, we have, 
\begin{equation}
D^{B,d}(x_F,\Lambda|s) = \kappa d^\pi_v(x^P)f_D(x^T|u_ss_s), 
\end{equation}
where $x^P\approx x_F$ and $x^T\approx m_\Lambda^2/(sx_F)$. 

Just as $pp\to \Lambda X$, there are also contributions 
from hyperon decays to $\pi^-p\to\Lambda X$. 
In $\pi^-p$ collisions, we have similar contributing processes 
from group (B) to $\Sigma^0$ and $\Xi^-$ production, i.e., 
$d_v^P+(u_ss_s)^T\to\Sigma^0$ 
and $d_v^P+(s_ss_s)^T\to\Xi^-$, respectively. 
Their corresponding number densities are given by, 
\begin{equation}
D^{B,d}(x_F,\Sigma^0|s) = \kappa d^\pi_v(x^P)f_D(x^T|u_ss_s), 
\end{equation}
\begin{equation}
D^{B,d}(x_F,\Xi^-|s) = \kappa d^\pi_v(x^P)f_D(x^T|s_ss_s), 
\end{equation}
respectively. 

Finally, we have, 
\begin{equation}
P_\Lambda(x_F|s)=\frac
{C \sum_{H_j=\Sigma^0,\Xi^-}t^D_{\Lambda,H_j} t^F_{H_j,d} D^{C,H_j}(x_F,\Lambda|s)}
{N_0(x_F,\Lambda|s)+D^{B,d}(x_F,\Lambda|s)
+\sum_{H_j=\Sigma^0,\Xi^-}D^{C,H_j}(x_F,\Lambda|s)},
\end{equation}
\begin{equation}
D^{C,H_j}(x_F,\Lambda|s)\approx \frac{m_{H_j}}{E_{\Lambda,H_j}}
D^{B,d}(\frac{m_{H_j}x_F}{E_{\Lambda, H_j}}, H_j|s),
\end{equation}
where the fragmentation spin transfer factors $t^F_{H_j,d}$'s 
for quark flavor $d$ and different hyperon $H_j$ are 
given in Table \ref{tab:1}, and the decay spin transfer factors 
$t^D_{\Lambda,H_j}$'s are given in Section III.

\begin{figure}
\includegraphics[]{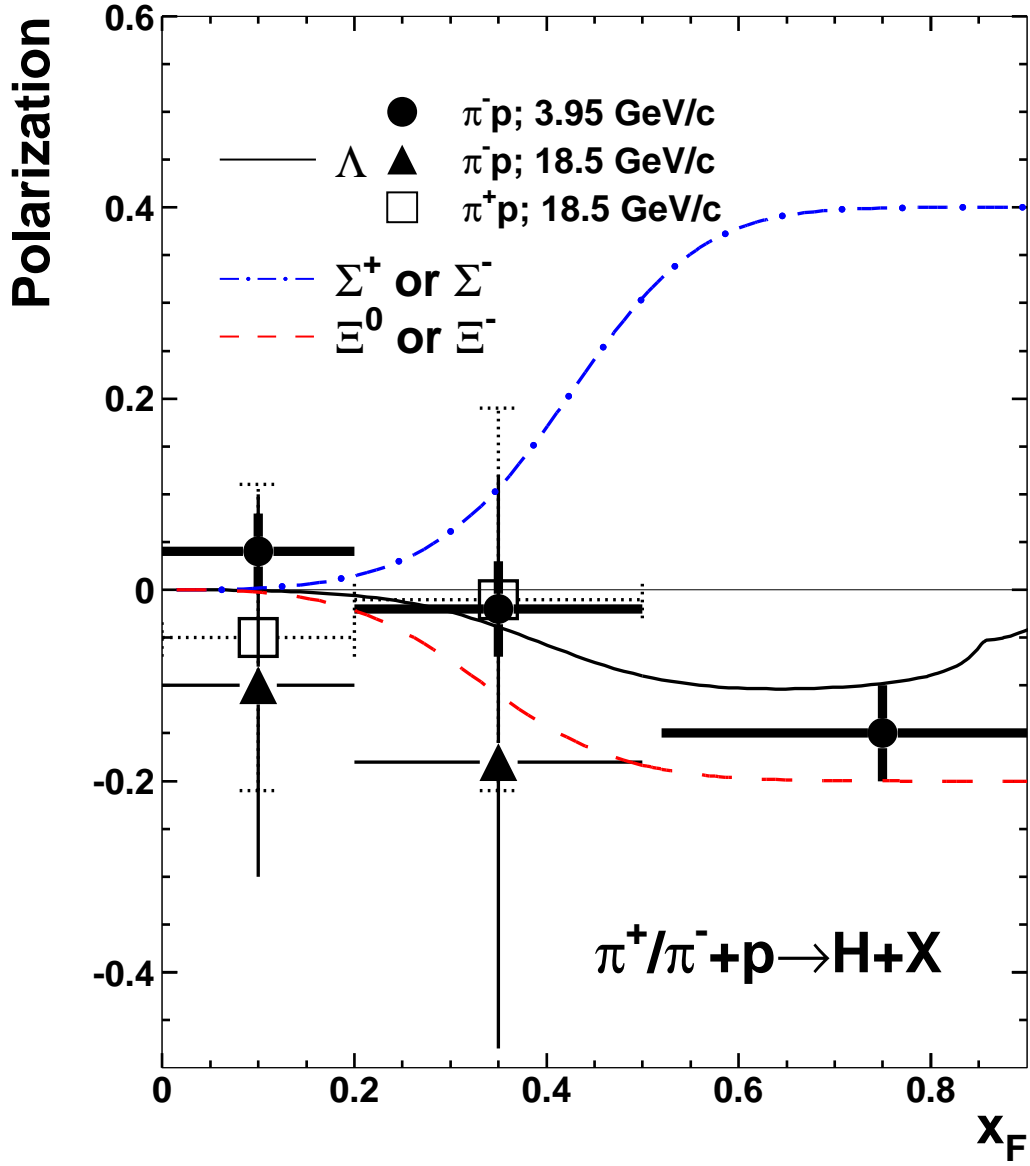}
\caption{\label{fig:pi-p2hyperon}
Calculated results for the polarizations of $\Lambda$, $\Sigma^{\pm}$ and $\Xi^{0,-}$ 
in $\pi p$ collisions 
as functions of $x_F$. 
Data for $\Lambda$ are taken from Refs. \cite{Ade84,Stu74}. }
\end{figure}

From Eq. (58), we see immediately that, 
since $t^F_{\Lambda,d}=0$, $P_\Lambda(x_F|s)$ in $\pi p\to\Lambda X$ 
should be equal to zero if the decay contributions are neglected. 
The nonzero $P_\Lambda$ in this process comes purely from the decays of 
heavier hyperons. 
Taking these decay contributions into account, 
we expect a small and negative $P_\Lambda(x_F|s)$ at large $x_F$ 
because $Ct^D_{\Lambda,\Sigma^0}t^F_{\Sigma^0,d}=-0.13$, 
and $Ct^D_{\Lambda,\Xi}t^F_{\Xi^-,d}=-0.18$. 
The numerical results obtained from Eq. (58) are given in Fig. \ref{fig:pi-p2hyperon}. 
In the calculations, we use the GRV-P LO set \cite{GRV92} for the unpolarized 
quark distribution function of pion 
and simply assume $N_0(x_F,H_i|s)$ in this process to be the same as 
that in $pp$ collisions. 
We clearly see that, as $x_F$ goes from 0 to 1, 
the obtained $P_\Lambda(x_F|s)$ starts from 0 
and decreases to about $-10\%$ at large $x_F$. 
These qualitative features are in agreement with the data \cite{Ade84,Stu74}.

\subsubsection{$\pi^-+p\to\Sigma^-$ (or $\Xi^-$)$+X$}
In $\pi^-p$ collisions, there is one contributing process from group (B) 
to the production of $\Sigma^-$ and $\Xi^-$. 
For $\Sigma^-$, it is $d_v^P+(d_ss_s)^T\to\Sigma^-$. 
The corresponding number density is given by,
\begin{equation}
D^{B,d}(x_F,\Sigma^-|s)=\kappa d^\pi_v(x^P)f_D(x^T|d_ss_s).
\end{equation}
The contributing process from (B) to $\Xi^-$ production 
has been given in the last subsection 
and its corresponding number density is given by Eq. (57). 
Thus, we have, 
\begin{equation}
P_{\Sigma^-}(x_F|s)=C\frac{t^F_{\Sigma^-,d}D^{B,d}(x_F,\Sigma^-|s)}
{N_0(x_F,\Sigma^-|s)+D^{B,d}(x_F,\Sigma^-|s)},
\end{equation}
\begin{equation}
P_{\Xi^-}(x_F|s)=C\frac{t^F_{\Xi^-,d}D^{B,d}(x_F,\Xi^-|s)}
{N_0(x_F,\Xi^-|s)+D^{B,d}(x_F,\Xi^-|s)}.
\end{equation}

As given in Table \ref{tab:1}, $t^F_{\Sigma^-,d}=2/3$, 
and $t^F_{\Xi^-,d}=-1/3$. 
We thus expect that $P_{\Sigma^-}(x_F|s)=0$ for $x_F$ near 0 
and increases to $Ct^F_{\Sigma^-,d}=0.4$ with increasing $x_F$. 
For $\Xi^-$, $P_{\Xi}(x_F|s)=0$ for $x_F$ near 0 
and decreases to $Ct^F_{\Xi^-,d}=-0.2$ with increasing $x_F$. 
This implies that $P_{\Sigma^-}(x_F|s)$ is positive in sign 
and increases fast to $40\%$ with increasing $x_F$, 
while $P_{\Xi^-}(x_F|s)$ is negative in sign 
and decreases to $-20\%$ with increasing $x_F$. 
The magnitude of $P_{\Sigma^-}$ should be larger than that of $P_{\Xi^-}$. 

Using these equations, 
We obtain the numerical results for 
$P_{\Sigma^-}(x_F|s)$ and $P_{\Xi^-}(x_F|s)$ in $\pi^-p$ collisions 
as shown in Fig. \ref{fig:pi-p2hyperon}. 
We see that the results show clearly the qualitative features mentioned above. 
These features can be tested by future experiments. 

\subsection{Hyperon polarization in $\Sigma^-+p\to H_i+X$}

It is interesting to note that experiments 
on hyperon polarizations in reactions using $\Sigma^-$ beam 
have also been carried out by the WA89 Collaboration at CERN. 
Some of the results have already been published \cite{WA89-95} 
and more results are coming. 
We now apply the proposed picture to this process. 

\subsubsection{$\Sigma^-+p\to\Lambda +X$}
Similar to $pp\to\Lambda X$, for $\Sigma^-p\to\Lambda X$, 
there are one contributing process from group (A), 
i.e., $(d_vs_v)^P+u_s^T\to\Lambda$ 
and two contributing processes from (B), i.e., 
$d_v^P+(u_ss_s)^T\to\Lambda$ and $s_v^P+(u_sd_s)^T\to\Lambda$.
Thus, we have, 
\begin{equation}
D^A(x_F,\Lambda|s) = \kappa_\Lambda^d f^{\Sigma^-}_D(x^P|d_vs_v)u_s(x^T), 
\end{equation}
\begin{equation}
D^{B,d}(x_F,\Lambda|s) = \kappa d^{\Sigma^-}_v(x^P)f_D(x^T|u_ss_s), 
\end{equation}
\begin{equation}
D^{B,s}(x_F,\Lambda|s) = \kappa s^{\Sigma^-}_v(x^P)f_D(x^T|u_sd_s). 
\end{equation}

There are also contributions from hyperon 
decays to $\Sigma^-p\to\Lambda X$. 
For $\Sigma^-p\to\Sigma^0(\mbox{or}\ \Xi^-)X$, we have similar contributing processes 
from groups (A) and (B) as those for $\Sigma^-p\to\Lambda X$. 
For $\Sigma^0$, they are 
$(d_vs_v)^P+u_s^T\to\Sigma^0$, 
$d_v^P+(u_ss_s)^T\to\Sigma^0$ and $s_v^P+(u_sd_s)^T\to\Sigma^0$.
The number densities of $\Sigma^0$ from these processes are given by, 
\begin{equation}
D^A(x_F,\Sigma^0|s) = \kappa_{\Sigma^0}^d f^{\Sigma^-}_D(x^P|d_vs_v)u_s(x^T), 
\end{equation}
\begin{equation}
D^{B,d}(x_F,\Sigma^0|s) = \kappa d^{\Sigma^-}_v(x^P)f_D(x^T|u_ss_s), 
\end{equation}
\begin{equation}
D^{B,s}(x_F,\Sigma^0|s) = \kappa s^{\Sigma^-}_v(x^P)f_D(x^T|u_sd_s), 
\end{equation}
respectively. 

For $\Xi^-$, they are 
$(d_vs_v)^P+s_s^T\to\Xi^-$, 
$d_v^P+(s_ss_s)^T\to\Xi^-$ and $s_v^P+(d_ss_s)^T\to\Xi^-$.
The corresponding number densities are given by, 
\begin{equation}
D^A(x_F,\Xi^-|s) = \kappa_{\Xi^-}^d f^{\Sigma^-}_D(x^P|d_vs_v)s_s(x^T), 
\end{equation}
\begin{equation}
D^{B,d}(x_F,\Xi^-|s) = \kappa d^{\Sigma^-}_v(x^P)f_D(x^T|s_ss_s), 
\end{equation}
\begin{equation}
D^{B,s}(x_F,\Xi^-|s) = \kappa s^{\Sigma^-}_v(x^P)f_D(x^T|d_ss_s), 
\end{equation}
respectively. 

For $\Sigma^-p\to\Xi^0X$, 
there is one contributing process from (B), i.e., 
$s_v^P+(u_ss_s)^T\to\Xi^0$. 
The corresponding number density is given by, 
\begin{equation}
D^{B,s}(x_F,\Xi^0|s) = \kappa s^{\Sigma^-}_v(x^P)f_D(x^T|u_ss_s). 
\end{equation}

Thus, we obtain that, 
\begin{equation}
N(x_F,\Lambda|s)=N_0(x_F,\Lambda|s)+D^A(x_F,\Lambda|s)
+\sum_{f=d,s}D^{B,f}(x_F,\Lambda|s)
+\sum_{H_j=\Sigma^0,\Xi^0,\Xi^-}D^{C,H_j}(x_F,\Lambda|s),
\end{equation}
\begin{equation}
D^{C,\Sigma^0}(x_F,\Lambda|s)\approx \frac{m_{\Sigma}}{E_{\Lambda,\Sigma^0}}
[D^A(\frac{m_{\Sigma}x_F}{E_{\Lambda,\Sigma^0}},\Sigma^0|s)
+\sum_{f=d,s}{D^{B,f}(\frac{m_{\Sigma}x_F}{E_{\Lambda,\Sigma^0}},\Sigma^0|s)}],
\end{equation}
\begin{equation}
D^{C,\Xi^0}(x_F,\Lambda|s)\approx \frac{m_{\Xi}}{E_{\Lambda,\Xi}}
D^{B,s}(\frac{m_{\Xi}x_F}{E_{\Lambda,\Xi}},\Xi^0|s), 
\end{equation}
\begin{equation}
D^{C,\Xi^-}(x_F,\Lambda|s)\approx \frac{m_{\Xi}}{E_{\Lambda,\Xi}}
[D^A(\frac{m_{\Xi}x_F}{E_{\Lambda,\Xi}},\Xi^-|s)
+\sum_{f=d,s}{D^{B,f}(\frac{m_{\Xi}x_F}{E_{\Lambda,\Xi}},\Xi^-|s)}]. 
\end{equation}

\begin{table}
\caption{\label{tab:4}
Relative weights $w(s_d,s_{dn}|d_vs_v)$ for 
the $d_vs_v$-diquark from $\Sigma^-$ in different spin states, 
those for the produced $\Lambda$ and $\Sigma^0$, 
the corresponding $\alpha^A_\Lambda$ and $\alpha^A_{\Sigma^0}$ 
and the total weights $w^A_\Lambda$ and $w^A_{\Sigma^0}$.}
\begin{tabular}{c|cc|cc|cc}
\hline 
Possible spin states
 & 
 \multicolumn{2}{c|}{$(d_vs_v)_{0,0}u_s^\downarrow$} & 
 \multicolumn{2}{c|}{$(d_vs_v)_{1,0}u_s^\downarrow$} & 
 \multicolumn{2}{c}{$(d_vs_v)_{1,1}u_s^\downarrow$} \\ \hline 
\begin{minipage}{4.2cm}
$w(s_d,s_{dn}|d_vs_v)$ 
\end{minipage}
 &
 \multicolumn{2}{c|}{$3/4$} & 
 \multicolumn{2}{c|}{$1/12$} & 
 \multicolumn{2}{c}{$1/6$} \\ \hline 
Possible products 
 & $\Lambda ^\downarrow $  & 
   $\Sigma^{0\downarrow }$ & 
   $\Lambda ^\downarrow $  & 
   $\Sigma^{0\downarrow }$ & 
   $\Lambda ^\uparrow $    & 
   $\Sigma^{0\uparrow }$ \\ \hline
$|\langle (q_vq_v)_{s_d,s_{dn}}q_s^\downarrow|H_i(s_n)\rangle|^2$
&  1/4  &  3/4 & 1/4 & $1/12$& 1/2 & $1/6$ \\ \hline
The final relative weights
 & 3/16  & 9/16 & 1/48 & $1/144$& 1/12 & $1/36$ \\ \hline
The resulted $w_{Hi}^A$ and $\alpha^A_{Hi}$ & 
 \multicolumn{3}{c}{$\Lambda$: $7/24$,  \ $-3/7$;} & 
 \multicolumn{3}{c}{$\Sigma^0$: $43/72$, \ $-39/43$} \\ \hline 
\end{tabular}
\end{table}

\begin{table}
\caption{\label{tab:5}
Relative weights $w(s_d,s_{dn}|d_vs_v)$ for 
the $d_vs_v$-diquark from $\Sigma^-$ in different spin states, 
those for the produced $\Xi^-$, 
the corresponding $\alpha^A_{\Xi^-}$ 
and the total weight $w^A_{\Xi^-}$.}
\begin{tabular}{c|c|c|c}
\hline 
Possible spin states
 & 
 $(d_vs_v)_{0,0}s_s^\downarrow$ & 
 $(d_vs_v)_{1,0}s_s^\downarrow$ & 
 $(d_vs_v)_{1,1}s_s^\downarrow$ \\ \hline 
\begin{minipage}{4.2cm}
$w(s_d,s_{dn}|d_vs_v)$ 
\end{minipage}
 &
 $3/4$ & 
 $1/12$ & 
 $1/6$ \\ \hline 
Possible products 
 & $\Xi^{-\downarrow }$ & 
   $\Xi^{-\downarrow }$ & 
   $\Xi^{-\uparrow }$ \\ \hline
$|\langle (q_vq_v)_{s_d,s_{dn}}q_s^\downarrow|H_i(s_n)\rangle|^2$
&  3/4  &  1/12 & 1/6  \\ \hline
The final relative weights
 & 9/16  & 1/144 & 1/36  \\ \hline
The resulted $w_{Hi}^A$ and $\alpha^A_{Hi}$ & 
 \multicolumn{3}{c}{$\Xi^-$: $43/72$, \ $-39/43$} \\ \hline 
\end{tabular}
\end{table}

In the same way as that of $pp\to \Lambda X$, 
we calculate $w^A_\Lambda$, 
the total relative weight for the production of 
$\Lambda$ from group (A), 
and $\alpha_\Lambda^A$. 
The results are shown in Table \ref{tab:4}. 
Similarly, we obtain the corresponding results for $\Sigma^0$ production
as shown in the same table 
and those for $\Xi^-$ production as shown in Table \ref{tab:5}. 
From these results, we obtain,
\begin{equation}
\Delta N(x_F,\Lambda|s)=C \alpha^A_\Lambda D^A(x_F,\Lambda|s)
+C t^F_{\Lambda,s} D^{B,s}(x_F,\Lambda|s)
+\sum_{H_j=\Sigma^0,\Xi^0,\Xi^-}\Delta D^{C,H_j}(x_F,\Lambda|s),
\end{equation}
\begin{equation}
\Delta D^{C,\Sigma^0}(x_F,\Lambda|s)\approx
C t^D_{\Lambda,\Sigma^0} \frac{m_{\Sigma}}{E_{\Lambda,\Sigma^0}}
[\alpha^A_{\Sigma^0} D^A(\frac{m_{\Sigma}x_F}{E_{\Lambda,\Sigma^0}},\Sigma^0|s)
+\sum_{f=d,s}{t^F_{\Sigma^0,f} D^{B,f}(\frac{m_{\Sigma}x_F}{E_{\Lambda,\Sigma^0}},
\Sigma^0|s)}],
\end{equation}
\begin{equation}
\Delta D^{C,\Xi^0}(x_F,\Lambda|s)\approx
C t^D_{\Lambda,\Xi} t^F_{\Xi^0,s} \frac{m_{\Xi}}{E_{\Lambda,\Xi}}
D^{B,s}(\frac{m_{\Xi}x_F}{E_{\Lambda,\Xi}},\Xi^0|s),
\end{equation}
\begin{equation}
\Delta D^{C,\Xi^-}(x_F,\Lambda|s)\approx
C t^D_{\Lambda,\Xi} \frac{m_{\Xi}}{E_{\Lambda,\Xi}}
[\alpha^A_{\Xi^-} D^A(\frac{m_{\Xi}x_F}{E_{\Lambda,\Xi}},\Xi^-|s)
+\sum_{f=d,s}{t^F_{\Xi^-,f} D^{B,f}(\frac{m_{\Xi}x_F}{E_{\Lambda,\Xi}},\Xi^-|s)}],
\end{equation}
where the fragmentation spin transfer factors $t^F_{H_j,f}$'s 
for different quark flavor $f$ and different hyperon $H_j$ are 
given in Table \ref{tab:1}, and the decay spin transfer factors 
$t^D_{\Lambda,H_j}$'s are given in Section III.

Using the results given by Eqs. (63--80), we can now calculate 
$P_\Lambda(x_F|s)$ in $\Sigma^-p$ collisions as a function of $x_F$. 
Between the two contributing processes from (B), 
the contribution from $s_v^P+(u_sd_s)^T\to\Lambda$ is 
more important than that from $d_v^P+(u_ss_s)^T\to\Lambda$ 
due to the strangeness suppression in the sea quarks of the target. 
We note that $\alpha^A_\Lambda=-3/7$, $t^F_{\Lambda,d}=0$, and $t^F_{\Lambda,s}=1$. 
Hence, if we neglect the decay contributions, 
we expect that, for $x_F$ increasing from 0 to 1, 
$P_\Lambda(x_F|s)$ starts from 0, increases to some positive value, 
then decreases to $C\alpha^A_\Lambda=-3C/7$ at $x_F\to 1$. 
We also note that $\alpha^A_{\Sigma^0}=-39/43$ 
which is large in magnitude and has the same sign as $\alpha^A_\Lambda$. 
This is quite different from the situation in $pp\to\Lambda X$ 
where $\alpha^A_{\Sigma^0}$ has different sign 
and smaller magnitude compared to $\alpha^A_\Lambda$. 
Since the decay spin transfer factor $t^D_{\Lambda,\Sigma^0}=-1/3$, 
we thus expect a very significant contribution from the $\Sigma^0$ decay 
to $P_\Lambda$ in $\Sigma^-p\to\Lambda X$, 
and this contribution cancels that from the directly produced $\Lambda$ 
at large $x_F$. 
As a consequence, 
taking the decay contributions into account, 
in particular the significant contribution from $\Sigma^0\to\Lambda\gamma$, 
we expect that, for $x_F$ going from 0 to 1, 
$P_\Lambda(x_F|s)$ starts from 0, increases slowly to some positive value 
at moderate $x_F$, 
and then begins to decrease at some $x_F$  
and finally reaches some negative value 
but the magnitude less than $3C/7$ at $x_F\to 1$. 

To get some quantitative feeling, 
we now use Eqs. (63--80) to calculate 
$P_\Lambda(x_F|s)$ in $\Sigma^-p$ collisions numerically. 
Since our purpose is to get a feeling of the $x_F$-dependence of $P_\Lambda(x_F|s)$, 
we simply make the following simplifications. 
For the unpolarized quark distribution functions in $\Sigma^-$, 
we approximately use those in proton \cite{GRV98} to replace them. 
For $N_0(x_F,H_i|s)$, we take it as the same as that in $pp$ collisions. 
The obtained numerical results for $P_\Lambda(x_F|s)$ 
are given in Fig. \ref{fig:Sigma-p2hyperon}. 
We see clearly that, as $x_F$ increases from 0 to 1, 
the obtained $P_\Lambda(x_F|s)$ first increases from 0 to some positive value, 
then decreases even to negative at $x_F\to 1$. 
These qualitative features are just those expected above. 
They can be checked by further experiments.

\subsubsection{$\Sigma^-+p\to\Sigma^{\pm}$ (or $\Xi^{0,-}$)$+X$}

\begin{table}
\caption{\label{tab:6}
Relative weights $w(s_d,s_{dn}|d_vs_v)$ for 
the $d_vs_v$-diquark from $\Sigma^-$ in different spin states, 
those for the produced $\Sigma^-$, 
the corresponding $\alpha^{A,ds}_{\Sigma^-}$ 
and the total weight $w^{A,ds}_{\Sigma^-}$.}
\begin{tabular}{c|c|c|c}
\hline 
Possible spin states
 & 
 $(d_vs_v)_{0,0}d_s^\downarrow$ & 
 $(d_vs_v)_{1,0}d_s^\downarrow$ & 
 $(d_vs_v)_{1,1}d_s^\downarrow$ \\ \hline 
\begin{minipage}{4.2cm}
$w(s_d,s_{dn}|d_vs_v)$ 
\end{minipage}
 &
 $3/4$ & 
 $1/12$ & 
 $1/6$ \\ \hline 
Possible products 
 & $\Sigma^{-\downarrow }$ & 
   $\Sigma^{-\downarrow }$ & 
   $\Sigma^{-\uparrow }$ \\ \hline
$|\langle (q_vq_v)_{s_d,s_{dn}}q_s^\downarrow|H_i(s_n)\rangle|^2$
&  3/4  &  1/12 & 1/6  \\ \hline
The final relative weights
 & 9/16  & 1/144 & 1/36  \\ \hline
The resulted $w_{Hi}^A$ and $\alpha^A_{Hi}$ & 
 \multicolumn{3}{c}{$\Sigma^-$: $43/72$, \ $-39/43$} \\ \hline 
\end{tabular}
\end{table}

\begin{table}
\caption{\label{tab:7}
Relative weights $w(s_d,s_{dn}|d_vd_v)$ for 
the $d_vd_v$-diquark from $\Sigma^-$ in different spin states, 
those for the produced $\Sigma^-$, 
the corresponding $\alpha^{A,dd}_{\Sigma^-}$ 
and the total weight $w^{A,dd}_{\Sigma^-}$.}
\begin{tabular}{c|c|c|c}
\hline 
Possible spin states
 & 
 $(d_vd_v)_{0,0}s_s^\downarrow$ & 
 $(d_vd_v)_{1,0}s_s^\downarrow$ & 
 $(d_vd_v)_{1,1}s_s^\downarrow$ \\ \hline 
\begin{minipage}{4.2cm}
$w(s_d,s_{dn}|d_vd_v)$ 
\end{minipage}
 &
 0 & 
 $1/3$ & 
 $2/3$ \\ \hline 
Possible products 
 & $\Sigma^{-\downarrow }$ & 
   $\Sigma^{-\downarrow }$ & 
   $\Sigma^{-\uparrow }$ \\ \hline
$|\langle (q_vq_v)_{s_d,s_{dn}}q_s^\downarrow|H_i(s_n)\rangle|^2$
&  ---  &  1/3 & 2/3  \\ \hline
The final relative weights
 & ---  & 1/9 & 4/9  \\ \hline
The resulted $w_{Hi}^A$ and $\alpha^A_{Hi}$ & 
 \multicolumn{3}{c}{$\Sigma^-$: 5/9, \ 3/5} \\ \hline 
\end{tabular}
\end{table}

First, we look at $\Sigma^-$ production in $\Sigma^-p$ collisions. 
For $\Sigma^-p\to\Sigma^-X$, there are two contributing processes from group (A), i.e., 
$(d_vs_v)^P+d_s^T\to\Sigma^-$ and $(d_vd_v)^P+s_s^T\to\Sigma^-$ 
and two from (B), i.e., 
$d_v^P+(d_ss_s)^T\to\Sigma^-$ and $s_v^P+(d_sd_s)^T\to\Sigma^-$.
The corresponding number densities are given by,
\begin{equation}
D^{A,ds}(x_F,\Sigma^-|s)=\kappa^d_{\Sigma^-}f^{\Sigma^-}_D(x^P|d_vs_v)d_s(x^T),
\end{equation}
\begin{equation}
D^{A,dd}(x_F,\Sigma^-|s)=\kappa^d_{\Sigma^-}f^{\Sigma^-}_D(x^P|d_vd_v)s_s(x^T),
\end{equation}
\begin{equation}
D^{B,d}(x_F,\Sigma^-|s)=\kappa d^{\Sigma^-}_v(x^P)f_D(x^T|d_ss_s),
\end{equation}
\begin{equation}
D^{B,s}(x_F,\Sigma^-|s)=\kappa s^{\Sigma^-}_v(x^P)f_D(x^T|d_sd_s).
\end{equation}
Hence, we have,
\begin{equation}
P_{\Sigma^-}(x_F|s)=C\frac
{\sum_{f=ds,dd}{\alpha^{A,f}_{\Sigma^-}D^{A,f}(x_F,\Sigma^-|s)}
+\sum_{f=d,s}{t^F_{\Sigma^-,f}D^{B,f}(x_F,\Sigma^-|s)}}
{N_0(x_F,\Sigma^-|s)
+\sum_{f=ds,dd}{D^{A,f}(x_F,\Sigma^-|s)}
+\sum_{f=d,s}{D^{B,f}(x_F,\Sigma^-|s)}},
\end{equation}
where $t^F_{\Sigma^-,f}$'s are given in Table \ref{tab:1}, 
$\alpha^{A,f}_{\Sigma^-}$ and other related quantities 
describing $\Sigma^-$ from the two contributing processes from (A) 
are shown in Table \ref{tab:6} and \ref{tab:7}, respectively. 

Because of the strangeness suppression in the sea quarks of the target, 
we expect that the most important contributing process 
to the production of $\Sigma^-$ from (A) is $(d_vs_v)^P+d_s^T\to\Sigma^-$ 
and that from (B) is $s_v^P+(d_sd_s)^T\to\Sigma^-$. 
From Tables \ref{tab:6} and \ref{tab:1}, we see that 
$\alpha^{A,ds}_{\Sigma^-}=-39/43$, and $t^F_{\Sigma^-,s}=-1/3$. 
Both of them contribute negatively to the polarization of $\Sigma^-$ 
in $\Sigma^-p\to\Sigma^- X$. 
In the contrary, from Tables \ref{tab:7} and \ref{tab:1}, we see that 
$\alpha^{A,dd}_{\Sigma^-}=3/5$, and $t^F_{\Sigma^-,d}=2/3$. 
Both of them are positive and $\alpha^{A,dd}_{\Sigma^-}<|\alpha^{A,ds}_{\Sigma^-}|$, 
but $t^F_{\Sigma^-,d}=2|t^F_{\Sigma^-,s}|$. 
We thus expect that a large portion of the contribution 
from $s_v^P+(d_sd_s)^T\to\Sigma^-$ to $P_{\Sigma^-}$ 
is cancelled by that from $d_v^P+(d_ss_s)^T\to\Sigma^-$, 
while a relatively small fraction of the contribution 
from $(d_vs_v)^P+d_s^T\to\Sigma^-$ 
is cancelled by that from $(d_vd_v)^P+s_s^T\to\Sigma^-$. 
Hence, for $x_F$ from 0 to 1, 
$P_{\Sigma^-}(x_F|s)$ in $\Sigma^-p\to\Sigma^- X$ 
should start from 0, 
decrease slowly to some negative value at moderate $x_F$, 
and then tend to a result 
between $C(\alpha^{A,ds}_{\Sigma^-}+\alpha^{A,dd}_{\Sigma^-})=-0.18$ 
and $C \alpha^{A,ds}_{\Sigma^-}=-0.54$ at $x_F\to 1$. 

\begin{figure}
\includegraphics[]{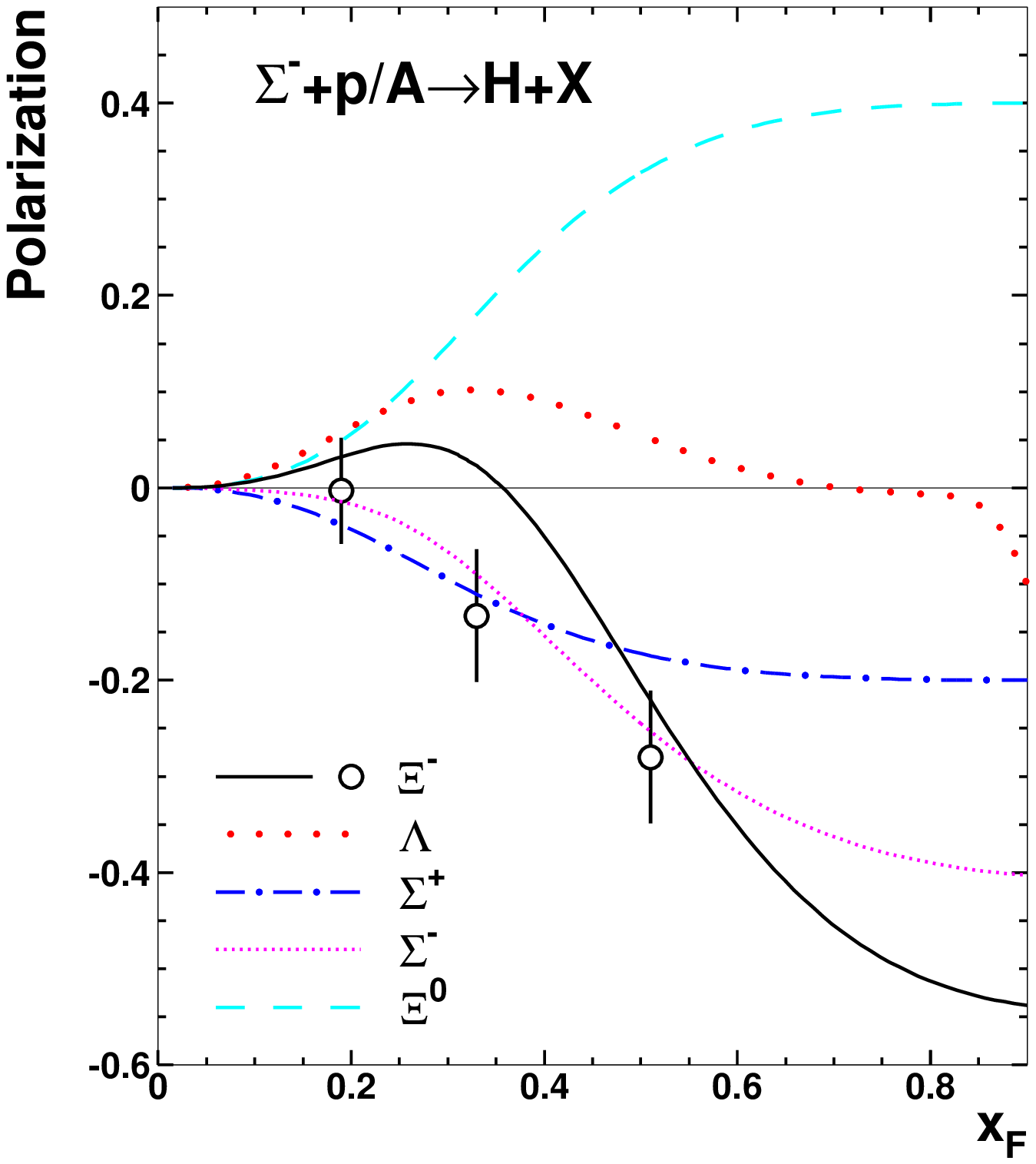}
\caption{\label{fig:Sigma-p2hyperon}
Calculated results for the polarizations of $\Lambda$, $\Sigma^{\pm}$ and $\Xi^{0,-}$ 
as functions of $x_F$. 
The thin dotted curve represents the polarization of $\Lambda$. 
Data for $\Xi^-$ are taken from Ref. \cite{WA89-95}. }
\end{figure}

Using Eq. (85), we calculate $P_{\Sigma^-}(x_F|s)$ in this process. 
The numerical results for it are given in Fig. \ref{fig:Sigma-p2hyperon}. 
We see that $P_{\Sigma^-}(x_F|s)$ indeed starts from 0, 
decreases to about $-20\%$ at moderate $x_F$ and reaches about $-40\%$ at $x_F\to 1$. 
These features are the same as those from the qualitative analysis 
and can be tested by future experiments. 

Then, we look at the production of $\Sigma^+$, $\Xi^0$ and $\Xi^-$ 
in $\Sigma^-p$ collisions. 
For $\Sigma^+$, there is no contributing process from group (A) but 
one contributing process from (B), i.e., 
$s_v^P+(u_su_s)^T\to\Sigma^+ $. 
The corresponding number density is given by, 
\begin{equation}
D^{B,s}(x_F,\Sigma^+|s)=\kappa s^{\Sigma^-}_v(x^P)f_D(x^T|u_su_s).
\end{equation}
The contributing processes to $\Xi$ production have been given 
in the last subsection and their corresponding number densities are 
given by Eqs. (69--72). 
Hence, we have,
\begin{equation}
P_{\Sigma^+}(x_F|s)=C\frac{t^F_{\Sigma^+,s}D^{B,s}(x_F,\Sigma^+|s)}
{N_0(x_F,\Sigma^+|s)+D^{B,s}(x_F,\Sigma^+|s)},
\end{equation}
\begin{equation}
P_{\Xi^0}(x_F|s)=C\frac{t^F_{\Xi^0,s}D^{B,s}(x_F,\Xi^0|s)}
{N_0(x_F,\Xi^0|s)+D^{B,s}(x_F,\Xi^0|s)},
\end{equation}
\begin{equation}
P_{\Xi^-}(x_F|s)=C\frac
{\alpha^A_{\Xi^-}D^A(x_F,\Xi^-|s)
+\sum_{f=d,s}{t^F_{\Xi^-,f}D^{B,f}(x_F,\Xi^-|s)}}
{N_0(x_F,\Xi^-|s)
+D^A(x_F,\Xi^-|s)
+\sum_{f=d,s}{D^{B,f}(x_F,\Xi^-|s)}},
\end{equation}
where the $t^F_{H_i,f}$'s are given in Table \ref{tab:1} 
and $\alpha^A_{\Xi^-}$ is given in Table \ref{tab:5}. 

We recall that $t^F_{\Sigma^+,s}=-1/3$, and $t^F_{\Xi^0,s}=2/3$. 
Thus, $P_{\Sigma^+}(x_F|s)$ should start from 0 and decrease to 
$Ct^F_{\Sigma^+,s}=-0.2$ with increasing $x_F$ 
while $P_{\Xi^0}(x_F|s)$ begin from 0 and increase to $Ct^F_{\Xi^0,s}=0.4$. 
This means that $P_{\Sigma^+}$ in this process is negative in sign 
and relatively small in magnitude but $P_{\Xi^0}$ positive and large. 
For $\Xi^-$ production, we have 
$\alpha^A_{\Xi^-}=-39/43$, $t^F_{\Xi^-,s}=2/3$, and $t^F_{\Xi^-,d}=-1/3$. 
Hence, we expect from Eq. (89) that, as $x_F$ increases from 0 to 1, 
$P_{\Xi^-}(x_F|s)$ starts from 0, 
increases to some positive value below $C t^F_{\Xi^-,s}=0.4$, 
begins to decrease at some $x_F$, and finally reaches to $C \alpha^A_{\Xi^-}=-0.54$. 
In Fig. \ref{fig:Sigma-p2hyperon}, we show the obtained numerical results 
as functions of $x_F$. 
We see clearly that the results indeed show the above-mentioned qualitative features. 
These features can be checked by future experiments. 

We emphasize again that the most important purpose of the numerical results 
presented in this section 
is to show the qualtitative features of hyperon polarizations 
in different reactions obtained in the proposed picture. 
For this purpose, we make several simplifications 
to reduce the free parameters in connection with the number densities 
in unpolarized reactions. 
Further improvements can be made if more accurate data are available. 

\section{summary and outlook}
\label{sec:summary}
In summary, we have calculated the polarizations of different hyperons 
as functions of $x_F$ in the inclusive $pp$, $K^-p$, $\pi^{\pm}p$, 
and $\Sigma^-p$ collisions. 
We used the picture proposed in a previous Letter \cite{LB97}, 
which relates the hyperon polarization 
in unpolarized hadron-hadron collisions 
to the left-right asymmetry in singly polarized reactions. 
We discussed the qualitative features 
for hyperon polarizations in these reactions 
and presented the corresponding numerical results. 
These qualitative features are all in agreement with the available data. 
Predictions for future experiments have been made. 
These predictions can be used as good tests to the picture. 

It should be emphasized that several points need to be further developed 
in the model. 
One of them is the influence of the vector meson production associated 
with the hyperon that contains a valence diquark of the projectile 
as mentioned at the end of Sec. \ref{sec:picture}. 
Such a study is under way. 
Another very important aspect is the transverse momentum dependence 
of the hyperon polarization. 
It is clear that the general formulae given in Sec. \ref{sec:method} 
can be extended to include the $p_T$-dependence. 
We can use them to calculate the $p_T$-dependence of $P_H$ 
in the model. 
This is also a very important aspect in the existing data 
and should be taken as a further challenge to the model. 
As can be seen from the formulae in Sec. \ref{sec:method}, 
the $p_T$-dependence of $P_H$ should come from the $p_T$-dependence of $C$,
and the interplay of the $p_T$-dependent $N_0$ and $D$. 
The $p_T$-dependance of the direct formation contribution $D$ 
comes mainly from those of the quark distributions. 
The $N_0$ part can be parameterized from the data 
on unpolarized cross sections. 
But, to get the $p_T$-dependence of $C$, 
we need the data on the $p_T$-dependence of $A_N$. 
Presently, there is no such data available. 
We get only an average $C$ in the $p_T$ interval $0.7<p_T<2.0$ GeV/$c$. 
This is the largest difficulty to calculate the $p_T$-dependence 
of $P_H$ at the present stage. 
Nevertheless, a phenomenological analysis can and should be made. 
Such studies are under way. 
The results we obtained in Sec. \ref{sec:results} should be taken 
as the average results in the corresponding $p_T$-region. 
Since many of the data are from fixed angle experiments, 
the comparison of our results with the data can only 
be regarded as qualitative. 

\begin{acknowledgments}
We thank Xu Qing-hua and other members of the theoretical particle physics group 
of Shandong University for helpful discussions. 
This work was supported in part by the National Science Foundation
of China (NSFC) and the Education Ministry of China. 
\end{acknowledgments}

\begin{thebibliography}{99}
\bibitem{ft:kappa} In practise, since there is no data
 directly on $N(x_F,\Lambda|s)$, we determine the two $\kappa$'s
 by fitting a few data points at large $x_F$ of
 the available data on the invariant cress section
 $Ed^3\sigma/dp^3$ at a given $p_T$
 because apparently $Ed^3\sigma/dp^3\propto x_FN(x_F,s)$.
 This implies that, we multiply the number densities
 $N(x_F,\Lambda|s)$, $N_0(x_F,\Lambda|s)$ and $D(x_F,\Lambda|s)$'s 
 with $x_F$ and a constant which changes
 the differential cross sections to the number densities.
 In this way, all of them are changed to their counterparts 
 in the invariant cross section $Ed^3\sigma/dp^3$.
 It is expected that, if the direct fusion model works, 
 $\kappa$ should neither strongly depend on $x$ nor on $p_T$. 
 The $x_F$ and $p_T$ dependences of the number density of the hadron 
 produced from this process come predominately from 
 the quarks and/or antiquarks in the initial states. 
 So we can approximately fix the $\kappa$ parameters 
 entering a $p_T$ integrated density from data at fixed $p_T$. 
 For $p_T =0.65$ GeV/$c$, 
 the corresponding result for $x_FN_0$ is 
 $0.2(1-x_F)^6e^{-3x_F^5}$ mb/GeV$^2$,
 that for $\kappa^d$ is $1.157\times 10^{-2}$ mb/GeV$^2$, 
 and that for $\kappa$ is $1.62\times 10^{-4}$ mb/GeV$^2$, 
 For more details, see Refs. \cite{BLM93,BL96,LB00rev} given above.
\end{thebibliography}
\end{document}